\documentclass{article}

\usepackage{amsmath, amssymb, amsthm, mathrsfs, bm}
\usepackage{geometry, extarrows, graphicx, enumitem, booktabs, subfigure}
\usepackage[round]{natbib}
\usepackage{titlesec}
\usepackage[hidelinks]{hyperref}
\usepackage{array}
\usepackage[dvipsnames]{xcolor}

\DeclareMathOperator*{\argmin}{argmin}

\def\bbE{\mathbb{E}}
\def\bbR{\mathbb{R}}

\def\cP{\mathcal{P}}

\theoremstyle{plain}
\newtheorem{theorem}{Theorem}
\newtheorem{proposition}[theorem]{Proposition}
\newtheorem{lemma}[theorem]{Lemma}
\newtheorem{corollary}[theorem]{Corollary}

\newtheorem{definition}{Definition}

\theoremstyle{remark}
\newtheorem{example}{Example}

\newtheorem{condition}{Scenario}

\newcommand{\revision}[1]{#1}

\begin{document}
\title{Invariant Probabilistic Prediction}
\date{\today}
\author{Alexander Henzi, Xinwei Shen, Michael Law, and Peter B\"uhlmann \\[0.5em] {\tt\small $\{$alexander.henzi, xinwei.shen, michael.law, peter.buehlmann$\}$@stat.math.ethz.ch}
\\[0.5em] Seminar for Statistics, \ ETH Z\"urich}
\maketitle

\begin{abstract}
In recent years, there has been a growing interest in statistical methods that exhibit robust performance under distribution changes between training and test data. While most of the related research focuses on point predictions with the squared error loss, this article turns the focus towards probabilistic predictions, which aim to comprehensively quantify the uncertainty of an outcome variable given covariates. Within a causality-inspired framework, we investigate the invariance and robustness of probabilistic predictions with respect to proper scoring rules. We show that arbitrary distribution shifts do not, in general, admit invariant and robust probabilistic predictions, in contrast to the setting of point prediction.  We illustrate how to choose evaluation metrics and restrict the class of distribution shifts to allow for identifiability and invariance in the prototypical Gaussian heteroscedastic linear model. Motivated by these findings, we propose a method to yield invariant probabilistic predictions, called IPP, and study the consistency of the underlying parameters. Finally, we demonstrate the empirical performance of our proposed procedure on simulated as well as on single-cell data.
\end{abstract}

\section{Introduction}
We study the problem of making probabilistic predictions for a response variable $Y\in\bbR$ from a set of covariates $X\in\bbR^d$. A probabilistic prediction for $Y$ is a probability distribution on $\mathbb{R}$, which, compared to point predictions, fully quantifies the uncertainty about $Y$ given the information in $X$. This, in turn, simultaneously yields point predictions for the mean and all quantiles, as well as prediction intervals. 
Commonly, probabilistic predictions are obtained by estimating the conditional distribution of $Y$ given $X = x$ from the training data and methods for distributional regression have been widely studied in literature with explicit statistical guarantees; see \citet{Kneib2021} for a general review. {When the training and test data have the same distribution}, such predictions can yield reliable test performance. 
However, distribution shifts often occur in real applications, which violates the fundamental i.i.d.\ assumption{, thus bringing challenges to the existing theory for extant prediction methods}. For instance, \cite{Henzi2023} {observed diminishing accuracy of distributional regression predictions over time for patient length of stay in some intensive care units due to evolving treatments and organization.} {Moreover, \cite{Vannitzsem2020} highlighted the challenges of changes to observation systems in weather forecasting, the field from which probabilistic predictions originated.}

Therefore, it is crucial to guarantee the accuracy of a prediction model under potential distribution shifts. 
One formulation of {predictive accuracy in such cases is through the worst-case performance among a class of shifted distributions, which is known as distributional robustness; see for example \cite{meinshausen2018causality}.  More formally, for a class of distributions $\cP$, one seeks to minimize the risk in predicting $Y$ uniformly over all $P \in \cP$.}
{One way to obtain distributional robustness is through the invariance property \citep{peters2016causal}, meaning the predictions admit a constant risk under a class of distributions with respect to a certain evaluation metric; see also \cite{Christiansen2021} and \cite{Krueger2021}.} 

Most of the existing literature on prediction under distribution shifts focuses on point prediction in regression or classification; see, for example, \cite{arjovsky2019invariant}, \cite{buhlmann2020invariance}, and \cite{duchi2021learning}. One popular approach is distributionally robust optimization as proposed by \cite{sinha2017certifiable}, who consider the minimax optimization over distributions within a certain distance or divergence ball of the observed data distribution.  In another line of work that includes \cite{Rothenhausler2021}, \cite{Christiansen2021}, and \cite{Shen2023}, a causality-inspired framework is adopted to exploit the underlying data structure, thereby enhancing prediction stability when faced with distributional shifts. 

For general predictions beyond the mean, however, distribution shifts have {not yet received much attention}. \revision{\citet{Nguyen2020conditional} develop distributionally robust maximum likelihood estimation in parametric families, which allows probabilistic predictions, but they only apply their method to logistic and Poisson regression which are essentially mean prediction problems. \citet{Chen2018}, \citet{Qi2022}, and \citet{Nguyen2020conditional} propose distributionally robust variants of quantile regression that allow to estimate certain conditional quantiles and generate prediction intervals. However, \citet{Chen2018} only consider median regression, \citet{Qi2022} do not allow conditional heteroscedasticity, and \citet{Nguyen2020conditional} do not apply their quantile prediction methods empirically. 
Recently, there has been an increasing interest in robustness in the conformal prediction literature, where the goal is to construct robust prediction intervals for a fixed marginal coverage level, rather than estimating the full distributions; see, e.g., \citet{Tibshirani2019}, \citet{Cauchois2024}, and \citet{Ai2024}.}  To the best of our knowledge, the only work \revision{about robust probabilistic prediction in the causality literature} is \cite{Kook2022}, who propose distributional anchor regression for conditional transformation models as a generalization of anchor regression \citep{Rothenhausler2021}. However, the theoretical properties of the procedure in \cite{Kook2022} are not addressed nor understood.  

\revision{The methodologically closest work to ours is by \citet{Krueger2021}, who propose to reduce a model's sensitivity to distribution shifts by reducing differences in risk across training domains. They focus on point predictions and classification but not on uncertainty quantification. One of the variants of their estimators, V-REx, turns out to be a special case of our proposed estimator which is derived in a more general manner after formalizing the invariance property of probabilistic predictions. In addition, they do not analyze the statistical properties of their estimators, nor do they discuss the existence of invariant predictions --- which, as we show, is not guaranteed in general.}

\revision{Our work gives the first rigorous formulation about invariance in probabilistic prediction, which sheds light on the possibilities and limitations, and we advocate a procedure with theoretical guarantees.} As a first step towards this aim, we consider models where the response variable $Y$ depends on $X$ through a location-scale transformation; that is, $Y = \mu^*(X) + \sigma^*(X)\varepsilon_Y$ with a noise variable $\varepsilon_Y$. The additive noise model is a special case of this model without heteroscedasticity, i.e., $\sigma^*(\cdot)$ constant, under which the do-interventional mean $\mu^*(X)$, as a prediction for $Y$, achieves a constant and worst-case optimal squared-error risk under all interventions on $X$; see for example \citet[Section 3]{Christiansen2021}. We formalize the invariance and robustness of $\mu^*(X)$ for not only conditional mean prediction but also more general settings.
We then extend the ideas from point prediction to probabilistic prediction by primarily defining the desired prediction model $\pi^*_{y|x}$ as the distribution of $Y$ under the do-intervention of $\mathrm{do}(X = x)$. We describe the fundamental difficulties by proving that for any sufficiently general model class, there exists no loss function under which $\pi^*_{y|x}$ has an invariant risk under all interventions. However, it turns out that for restricted interventions and certain loss functions, invariance and robustness are possible in location-scale models, and we show that one can characterize $\pi^*_{y|x}$ through these two properties in the case of the Gaussian heteroscedastic linear model.

In addition to our fundamental derivations, we consider the problem of obtaining invariant probabilistic predictions in a setting where data from environments with different interventions on $X$ are available. We define the invariant probabilistic prediction (IPP) in terms of a concrete procedure. 
Its consistency is established for identifiable parametric models.
In addition, we propose a rule to select a suitable penalization parameter in finite samples.
Furthermore, we demonstrate the efficacy of IPP in a well-specified simulation setting with the Gaussian heteroscedastic linear model, \revision{as well as in a scenario
with misspecification} and a real application on single-cell data. \revision{In the empirical applications, IPP is competitive to or outperforms other methods that are specifically designed for deriving robust point predictions or prediction intervals.}

\section{Background and setup}\label{sec:prob_setup}

\subsection{Model for observational distribution}
We consider the general model
\begin{equation} \label{eq:model}
    X =\varepsilon_X, \quad Y \leftarrow f^*(X, \varepsilon_Y),
\end{equation}
{where the distribution of $X \in \bbR^d$ is arbitrary, and the response variable $Y$ is given by a structural equation. We use the notation $X = \varepsilon_X$ only because later we will have shifted distributions for $X$. The function $f^*(\cdot,\cdot)$ depends in its first argument only on a subset $S^*$ of the covariates $X$; namely, $S^*$ equals the set of the parental variables of $Y$ which are among the covariates $X$; mathematically,  $f^*(x,\varepsilon_Y)= f^*(x_{S^*},\varepsilon_Y)$ for all $x \in \bbR^{d}$. 
The noise variables $\varepsilon_Y$ may be dependent of $X = \varepsilon_X$ due to hidden confounding between $Y$ and $X$.}  
The relationship between $X$ and $Y$ is very general, allowing causal and anti-causal directions. 
We denote the joint distribution \revision{$\mathcal{L}(X,Y)$} by $P^o$ and the conditional distribution \revision{$\mathcal{L}(Y|X=x)$} by $P^o_{y|x}$.  

\begin{figure}
    \centering
    \includegraphics[width=\textwidth]{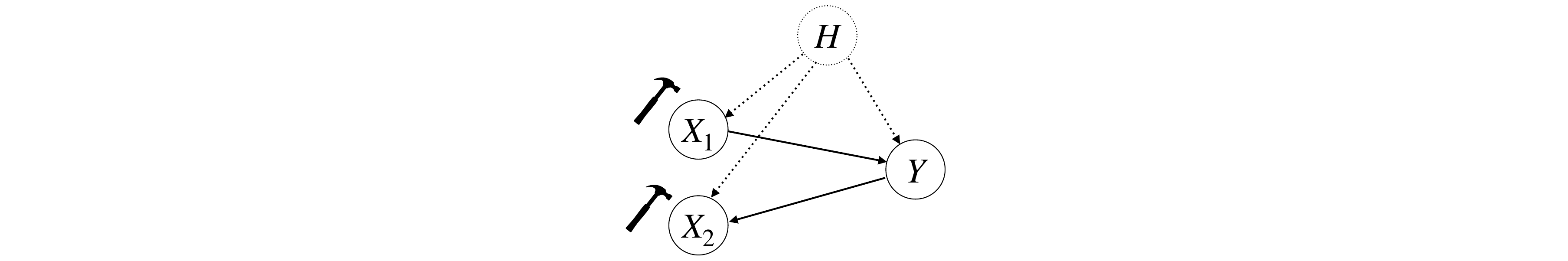}\vspace{-0.1in}
    \caption{Graphical structure of an example of models \eqref{eq:model} {and \eqref{eq:perturbation}. There is a hidden confounding variable $H$ and the hammers indicate that perturbations or interventions happen at $X = (X_1,X_2)$ as in model \eqref{eq:perturbation}.}}
    \label{fig:graph}
\end{figure}

{\subsection{Model for interventional distributions and heterogeneous data}}

In practical applications, interventions often happen to the system, leading to heterogeneity in the data. 
We consider here interventions on the covariates $X$. {Our model for such interventional distributions is as follows. In model \eqref{eq:model}, the random variable $X$ is replaced by the a new variable transformed with a function $g$,
\begin{equation}\label{eq:perturbation}
\revision{X^{\mathrm{int}}} = g(\varepsilon_X), \qquad
Y^{\mathrm{int}} \leftarrow f^*(X^{\mathrm{int}},\varepsilon_Y). 
\end{equation}
Here, the function $g(\cdot)$ could be random or deterministic and its argument is $\varepsilon_X$. This formulation encompasses various perturbation models. For example, if $g(\varepsilon_X) \equiv x$ for a deterministic value $x$, this is a hard intervention and equals the do-operator $\mathrm{do}(X = x)$, as formally defined below. If $g(\varepsilon_X) = \varepsilon_X + U$ for some random $U$, this is a shift intervention with random shift $U$. Thus, our perturbation model in \eqref{eq:perturbation} is rather general, \revision{and, as illustrated in examples in the article and in Appendix \ref{sec:sim2}, allows interventions that simultaneously affect the marginal distribution of the covariates and the conditional distributions of the response variable given the covariates}. We illustrate models \eqref{eq:model}-\eqref{eq:perturbation} in~Fig.~\ref{fig:graph}.

\begin{definition}[Do-interventional distribution]
In the perturbation model \eqref{eq:perturbation}, let $X^{\mathrm{int}}=x$ for any $x \in \mathbb{R}^d$. We refer to the distribution of $Y^{\mathrm{int}}=f^*(x, \varepsilon_Y)$ as the do-interventional distribution of $Y$ under $\mathrm{do}(X=x)$, denoted by $\pi^*_{y|x}$.  
\end{definition}
We remark that $\pi^*_{y|x}$ is the object \revision{of interest}, while the model in \eqref{eq:perturbation}, which will be viewed later in \eqref{eq:environments} as the one for generating heterogeneous data, is more general than containing only do-interventions.

Consider now the setting where the data comes from different environments or sources $e=1,\ldots ,E$ \citep{peters2016causal,Rothenhausler2017CausalDF,arjovsky2019invariant}. We model the data as realizations for each environment $e$ as
\begin{equation}\label{eq:environments}
    X^e = g^e(\varepsilon_X),\ \ Y^e \leftarrow f^*(X^e,\varepsilon_Y),\ e=1,\ldots ,E.
\end{equation}
Thus, the model for each environment in \eqref{eq:environments} is an intervention model as in \eqref{eq:perturbation}, where each environment $e$ has its own perturbation or assignment function $g^e(\cdot)$.} 
\revision{We emphasize that 
our model for interventions and our theory only allow interventions on the covariates \revision{but not on latent variables. This is a standard assumption though in causality \citep{rothenhausler2015backshift,rothenhausler2019causal}. Our results in simulations and on real data suggest in addition} that predictions with IPP can be robust also under interventions on latent confounders.} 

\subsection{Proper scoring rules}
{To evaluate the accuracy of probabilistic predictions, we consider proper scoring rules.} A proper scoring rule {for a class of probability distributions on the real line}, is a function {$S : \mathcal{F} \times \mathbb{R} \to [-\infty, \infty]$} taking as arguments {$F \in \mathcal{F}$} and an observation {$Y \in \mathbb{R}$} {satisfying} 
\begin{equation} \label{eq:proper_score}
    \mathbb{E}_{Y\sim F}[S(F,Y)] \leq \mathbb{E}_{Y\sim F}[S(G,Y)]
\end{equation}
for all distributions $F, G \in \mathcal{F}$, where it is assumed that the expectations exist. {A proper} scoring rule $S$ is called strictly proper with respect to $\mathcal{F}$ if equality in {equation} \eqref{eq:proper_score} {holds only} if $F = G$. In Table \ref{tab:proper_scoring}, we give an overview of some commonly used proper scoring rules and we refer the interested reader to 
\citet{Gneiting2007} for a general review. 
{The two most popular proper scoring rules are the logarithmic score, which corresponds to negative log-likelihood, and the CRPS, which can be equivalently expressed as $\mathrm{CRPS}(F,Y) = \int_{-\infty}^{\infty} \{1(Y\leq z) - F(z)\}^2 \, dz$.
\begin{table}[t]
    \centering
    \small
    \begin{tabular}{cc}
        Logarithmic score (LogS) & Continuous ranked probability score (CRPS) \\[0.2em]
        
        $-\log\{g(Y)\}$ & $\mathbb{E}_{\eta \sim G}|Y-\eta| - \mathbb{E}_{\eta, \tilde{\eta} \sim G}|\eta-\tilde{\eta}|/2$ \\[1em]

        Scale invariant CRPS \citep[SCRPS;][]{Bolin2023} & Pseudospherical score (PseudoS) \\[0.2em]
        $\mathbb{E}_{\eta \sim G}|Y-\eta|/\mathbb{E}_{\eta, \tilde{\eta} \sim G}|\eta-\tilde{\eta}| + \log(\mathbb{E}_G|\eta-\tilde{\eta}|)/2$ & $-g(Y)^{\alpha-1} \{\int_{-\infty}^{\infty}g(z)^{\alpha})\,dz\}^{1/\alpha - 1}$ \\[1em]

        Quadratic score (QS) & Hyv\"arinen score \citep[HyvS;][]{Hyvarinen2005} \\
        $-2g(Y) + \int_{-\infty}^{\infty}g(z)^2\,dz$ & $2g''(Y)/g(Y) - \{g'(Y)/g(Y)\}^2 $\\[1em]
    \end{tabular}
    \caption{Examples of strictly proper scoring rules for outcome $Y \in \mathbb{R}$ and a probability prediction $G$ with density $g$. In CRPS and SCRPS, the expectation is over independent $\eta,\eta'$ with distribution $P$. For PseudoS, we have $\alpha > 1$. See \citet{Gneiting2007} for the scores without explicit reference.}
    \label{tab:proper_scoring}
\end{table}

{Analogous to proper scoring rules for probabilistic predictions, \citet{Murphy1985} introduced the idea of consistent scoring functions for point predictions; see also \citet{Gneiting2011}.  For a real-valued univariate functional $T : \mathcal{F} \to \mathbb{R}$, a scoring function $L: \mathbb{R} \times \mathbb{R} \to [0, \infty)$ is consistent if $\mathbb{E}_{Y\sim F} [L\{T(F), Y)\}] \leq \mathbb{E}_{Y\sim F} \{L(t, Y)\}$ for all $t \in \mathbb{R}$ and $F \in \mathcal{F}$; if equality holds in the above display only if $t = T(F)$, then $L$ is a strictly consistent scoring function.  As an example, the usual squared error loss $L(t, Y) = (t - Y)^{2}$ is a strictly consistent scoring function with respect to all distributions with finite second moment.  By Theorem 3 of \cite{Gneiting2011}, if $L$ is a consistent scoring function for $T$, then $S(F, Y) := L\{T(F), Y\}$ is a proper scoring rule, but not necessarily strictly proper.
}

In the standard definitions of both proper scoring rules and consistent scoring functions, the prediction is fixed. Typically, a prediction for an outcome variable $Y$ depends on observed covariates $X$, and in this case, the distributions and expected values in the definitions are to be understood conditional on $X$. We write $\pi_{y|x}$ to denote a probabilistic prediction for $Y$ given $X = x$. Formally, $\pi_{y|\cdot}$ is a Markov kernel, meaning that $\pi_{y|x}$ is a probability distribution for each fixed $x \in \mathbb{R}^d$, and $x \mapsto \pi_{y|x}(A)$ is measurable for each Borel set $A \subseteq \mathbb{R}$. The subscript $y$ in $\pi_{y|x}$ is only notation to highlight that $\pi_{y|x}$ is a prediction for the response variable $Y$ given $X=x$, and it has no mathematical implications; also, $\pi_{y|x}$ is generally not identical to the (observational) conditional distribution of $Y$ given $X = x$. We define the risk of a prediction with respect to a proper scoring rule as
\begin{align*}
    \mathcal{R}(\pi_{y|\cdot}, P, S) = \mathbb{E}_{(X,Y) \sim P}\{S(\pi_{y|X}, Y)\}.
\end{align*}

\revision{\subsection{An illustrative example} \label{sec:toy_example}
\begin{figure}[t]
    \centering
    \includegraphics[width=0.8\textwidth]{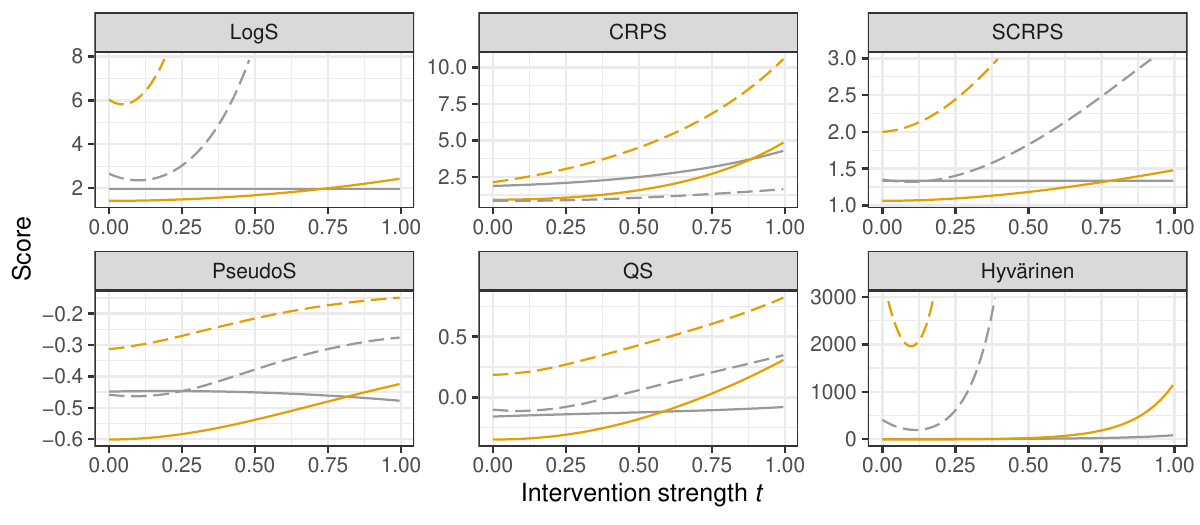}\vspace{-0.1in}
    \caption{Scores for the example from Section \ref{sec:restricted_interventions}, with $\alpha = 2$ for PseudoS. \revision{Solid} lines are for $X^t = \Gamma^t\varepsilon_X$, dashed lines for the do-interventions, in \revision{orange} color for $P^o_{y|x}$ and \revision{gray} for $\pi^*_{y|x}$. \label{fig:illustrative_example}}
\end{figure}
To make the definitions up to this point more specific, we present an illustrative example.  Consider the model 
\begin{align*}
    \begin{pmatrix}
  \varepsilon_Y \\
  \varepsilon_{X_1} \\
  \varepsilon_{X_2}
\end{pmatrix} \sim \mathcal{N}_3\left\{0, 
    \begin{pmatrix}
       3  & 1 & 1\\
       1  & 1 & 0\\
        1 & 0 & 1 \\
    \end{pmatrix}\right\}\!, \quad
        X= X = (\varepsilon_{X_1},\varepsilon_{X_2})^{\top}, \quad Y \leftarrow X_1
    + \exp(X_2)\varepsilon_Y
\end{align*}
with the test intervention classes $\{ \mathrm{do}(X_1^t = t, X_2^t = -1.5 + t^2) : 0 \leq t \leq 1\}$ and 
\begin{align} \label{eq:matmult_int}
    \{ X^t = \Gamma^t\varepsilon_X : \Gamma^t = \begin{pmatrix} 1 & t \\ t & 1 \end{pmatrix}, 0 \leq t \leq 1 \}.
\end{align}
Here, $t$ measures the strength of the intervention. Both interventions simultaneously affect the marginal distribution of $X$ and the mean of the distribution of $Y^t|X^t = x^t$, which is
\[
    \mathcal{N}\{x_1^t + (1+t)(x_1^t+x_2^t)\exp(2x_2^t + 2tx_1^t), \exp(2x_2^t)\}
\]
for the interventions \eqref{eq:matmult_int}. Figure \ref{fig:illustrative_example} plots the risk of the do-interventional distribution $\pi^*_{y|\cdot}$ and of the conditional distribution $P^o_{y|\cdot}$ as a function of the intervention strength $t$. Later in Section \ref{sec:restricted_interventions}, we prove that the LogS and SCRPS of the prediction $\pi^*_{y|\cdot}$ are constant under the intervention $X^t = \Gamma^t\varepsilon_X$ for all $t$. This is not true for the conditional distribution $P^o_{y|\cdot}$ in the observational environment, which is optimal for $t = 0$ under the interventions \eqref{eq:matmult_int} but deteriorates as $t$ increases. Furthermore, the prediction $\pi^*_{y|\cdot}$ is optimal under the do-intervention, since it equals the true conditional distribution, but as the figure shows, it does not admit an invariant risk.}

\section{Invariance and robustness in location-scale models}\label{sec:lsm}

\subsection{Impossibility of invariance under arbitrary interventions}\label{sec:inv_point_pred}
In this section, we investigate invariance and robustness for location-scale models, specializing model \eqref{eq:model} to
\begin{equation} \label{eq:location_scale}
    X = \varepsilon_X, \quad Y \leftarrow \mu^*(X) + \sigma^*(X)\varepsilon_Y,
\end{equation}
where $\mu^*\colon \mathbb{R}^d \rightarrow \mathbb{R}$ and $\sigma^*\colon \mathbb{R}^d \rightarrow (0,\revision{\infty})$ are the location and scale functions respectively, {and $\varepsilon_Y$ may be dependent on $X = \varepsilon_X$ as in model \eqref{eq:model} due to hidden confounding}. 
To motivate our definitions and results for probabilistic predictions, we first review the well-known robustness properties of the function $\mu^*(.)$ in point prediction without heteroscedasticy; that is, in the special case where $\sigma^*(X) \equiv 1$. Formally, a point prediction $\tilde{\mu}$ is invariant with respect to a class of distributions $\cP$ and a loss function $L$ if 
\begin{align}\label{eq:point:invariant}
    \mathbb{E}_{P}[L\{\tilde{\mu}(X), Y\}] = \mathbb{E}_{P'}[L\{\tilde{\mu}(X), Y\}], \quad P,P' \in \cP.
\end{align}
Similarly, an invariant point prediction $\tilde{\mu}$ is robust if 
\begin{align}\label{eq:point:robust}
    \mathbb{E}_{P}[L\{\tilde{\mu}(X), Y\}] = \inf_{\mu} \sup_{P \in \cP} \mathbb{E}_{P}[L\{{\mu}(X), Y\}]
\end{align}
where the infimum is over all measureable functions $\mu : \mathbb{R}^{d} \to \mathbb{R}$.  Results for point prediction are commonly formulated for mean prediction with squared error loss but, as discussed in \citet{Christiansen2021}, there is an interest to study robustness for other loss functions. To this end, the following proposition provides such an extension. \revision{Its proof, as well as the proofs of all other theoretical results, can be found in Appendix \ref{sec:proofs}.}

\begin{proposition} \label{prop:additive_error}
Consider the additive noise model
\begin{equation} \label{eq:additive_error_model}
    X = \varepsilon_X, \quad Y \leftarrow \mu^*(X) + \varepsilon_Y,
\end{equation}
as a special case of the model in \eqref{eq:location_scale}. 
Let $T$ be a functional for which $T\{\mathcal{L}(\varepsilon_Y)\} = 0$, and let $L(t,Y)$ be a strictly consistent scoring function for $T$ that depends on $t$ and $Y$ only through $Y-t$ and for which $\mathbb{E}\{L(0, \varepsilon_Y)\}$ exists. Let $\mathcal{P}$ be the set of all interventional distributions of $(X,Y)$ under the model given in \eqref{eq:additive_error_model}. Then, $\mu^{\ast}(\cdot)$  is the unique function satisfying \eqref{eq:point:invariant} and \eqref{eq:point:robust}.
\end{proposition}

An example where the above proposition extends mean prediction with the squared error loss is quantile prediction. If the $\alpha$-quantile of $\varepsilon_Y$ is unique and equal to zero, then $\mu^*(\cdot)$ gives a worst-case optimal prediction for the $\alpha$-quantile under the loss quantile loss $L(t, Y) = \{1(t - Y \geq 0) - \alpha\}(t - Y)$. Interestingly, to achieve robustness in mean prediction without further assumptions, we must use the squared error loss function as it is the only consistent scoring function for the mean that depends on $t$ and $Y$ merely through $Y - t$ \revision{\citep{Gneiting2011}}. 

We now generalize the ideas of invariance and robustness from point prediction to probabilistic prediction, which we define for the general model given in \eqref{eq:model}.

\begin{definition}[Invariance]\label{def:invariance}
	A probabilistic prediction $\pi_{y|\cdot}$ is said to be invariant {with respect to} a strictly proper scoring rule $S$ and a class of distributions $\cP$ if there exists a constant $c \in \mathbb{R}$ such that
	\begin{equation*}
		\mathcal{R}(\pi_{y|\cdot}, P, S)\equiv c,\ P\in\cP.
	\end{equation*} 
\end{definition}
{For convenience, we refer to such predictions as $(S, \cP)$- invariant.}

\begin{definition}[Robustness]\label{def:robust}
    A probabilistic prediction $\pi_{y|\cdot}$ is said to be distributionally robust {with respect to} a strictly proper scoring rule $S$ and a class of distributions $\cP$ if 
	\begin{equation*}
		\sup_{P\in\cP}\mathcal{R}(\pi_{y|\cdot}, P, S) \leq \sup_{P\in\cP}\mathcal{R}(\pi'_{y|\cdot}, P, S)
	\end{equation*} 
	for all probabilistic predictions $\pi'_{y|\cdot}$. We denote such predictions as $(S, \cP)$-robust.
\end{definition}

For the location-scale model, we have $\pi_{y|x}^{\ast} = \mathcal{L}\{\mu^{\ast}(x) + \sigma^{\ast}(x) \varepsilon_{Y}\}$ for any $x$; that is, $\{\pi_{y|x}^\ast\}$ are a location-scale family with baseline distribution $Q^{\ast}$.  This is analogous to the setting of point prediction \revision{with} the function $\mu^{\ast}(X) = T\{\mathcal{L}(\mu^{\ast}(X) + \varepsilon_{Y})\}$ under the do-intervention $\mathrm{do}(X = x)$. Then, by the same arguments as in Proposition \ref{prop:additive_error}, we have that invariance implies robustness in the case of probabilistic prediction, as shown in the following proposition.

\begin{proposition} \label{prop:invariance_probabilistic}
{Consider the general model given in \eqref{eq:model} and let $S$ be a strictly proper scoring rule and $\cP$ be the family of all interventional distributions.  If $\pi_{y|\cdot}^{\ast}$ is $(S, \cP)$-invariant, then $\pi_{y|\cdot}^{\ast}$ is $(S, \cP)$-robust and the risk under the observational distribution $P^o$, $\mathcal{R}(\pi^*_{y|\cdot}, P^o, S)$, is minimal among the risks of all invariant $(S,\cP)$-invariant predictions}.
\end{proposition}


{Compared to Proposition \ref{prop:additive_error}, which shows that there exists an invariant and robust prediction for suitable strictly consistent scoring functions, Proposition \ref{prop:invariance_probabilistic} only shows that if $\pi_{y|\cdot}^{\ast}$ is invariant, then it is also robust.  As the following theorem demonstrates, if the class of distributions $\cP$ contains all interventional distributions, there in fact does not necessarily exist any invariant prediction.}

\begin{theorem} \label{thm:score_not_exist}
Assume the model in \eqref{eq:location_scale} with
\begin{equation} \label{eq:counterexample_model}
    X = (X_1, X_2)^{\top} = (\varepsilon_{X_1}, \varepsilon_{X_2})^{\top}, \quad Y \leftarrow \mu^{\ast} (X_{1}) + \sigma^{\ast}(X_{2}) \varepsilon_{Y},
\end{equation}
{and additionally}
\begin{enumerate}
    \item {$\varepsilon_{Y} \sim \mathcal{N}(0,1)$ marginally};
    \item the functions $\mu^*\colon \mathbb{R} \rightarrow \mathbb{R}$ and $\sigma^* \colon \mathbb{R} \rightarrow (0,\infty)$ are surjective; {and}
    \item $\mathcal{P}$ contains the do-interventional distributions for all interventions on $X$ with any assignment value $x \in \bbR^2$.
\end{enumerate} 
Then there exists no strictly proper scoring rule $S$ such that $\pi^*_{y|\cdot}$ is $(S,\cP)$-invariant.
\end{theorem}

{The model in \eqref{eq:counterexample_model} is a Gaussian location-scale model, which is a special case of the general model given in \eqref{eq:model} \revision{and includes the common Gaussian heteroscedastic linear model (see Example 1 below)}.  Thus, Theorem \ref{thm:score_not_exist} shows that invariance in the sense of Definition \ref{def:invariance} is generally not possible when $\cP$ is the class of all interventional distributions on $X$. 
Hence, to obtain invariance, we must restrict our attention to finer classes of interventional distributions. }From a practical perspective, it often fits the real-world scenario better not to look for robustness against all interventions, since some may not be encountered and considering all of them may lead to overly conservative predictions.

\subsection{Restricted interventions} \label{sec:restricted_interventions}
Naturally, there always exists a class of interventional distributions for which the proper scoring rule admits an invariant prediction; for example, if we consider the class of all interventional distributions with $\sigma^\ast(X) = 1$, then we return to the setting of Proposition \ref{prop:additive_error}.  However, such an intervention class is not of particular interest.  As a first step towards identifying suitable and potentially practical interventional distributions, we explicitly compute the risk based on all six proper scoring rules given in Table \ref{tab:proper_scoring}.

\begin{lemma}
Let $Q^o$ be the marginal distribution of $\varepsilon_Y$.
{For the model given in \eqref{eq:location_scale}}, we have
\[
    \begin{array}{ll}
     \mathrm{CRPS}(\pi_{y|X}^*, Y)
    = \sigma^*(X)\mathrm{CRPS}(Q^o, \varepsilon_Y),
    & \mathrm{LogS}(\pi_{y|X}^*, Y) = \mathrm{LogS}(Q^o, \varepsilon_Y) + \log\{\sigma^*(X)\}, \\
     \mathrm{SCRPS}(\pi^*_{y|X}, Y) = \mathrm{SCRPS}(Q^o, \varepsilon_Y) + \log\{\sigma^*(X)\}/2,
    & \mathrm{PseudoS}(\pi_{y|X}^*, Y) = \sigma^*(X)^{1/\alpha - 1}\mathrm{PseudoS}(Q^o, \varepsilon_Y), \\
     \mathrm{QS}(\pi_{y|X}^*, Y) = \sigma^*(X)^{-1}\mathrm{QS}(Q^o, \varepsilon_Y),
    & \mathrm{HyvS}(\pi_{y|X}^*, Y) = \sigma^*(X)^{-2}\mathrm{HvyS}(Q^o, \varepsilon_Y),
    \end{array}
\]
where it is assumed that $Q^o$ admits a density with respect to Lebesgue measure for the cases of LogS, PseudoS, and QS, and that the density is twice differentiable for the case of HyvS.
\end{lemma}

The above result implies that, to formulate a restricted intervention class for invariant predictions under LogS and SCRPS, we only need knowledge of the marginal distribution of $X$.  On the other hand, for CRPS, PseudoS, QS, and HyvS, knowledge of the joint distribution between $X$ and $\varepsilon_{Y}$ is generally required.  In practice, the latter necessitates understanding of the underlying structural and confounding mechanism, which, in most applications, is not realistic.  Meanwhile, for LogS and SCRPS, an invariant risk is equivalent to $\mathbb{E}_P[\log\{\sigma^*(X)\}]$ being constant under interventional distributions $P$, which depends on the covariates and the interventions only and not on the confounding mechanism.  This leads us to the following result.

\begin{lemma}[Exponential scale parametrization] \label{lem:exponential_scale}
For the location-scale model given in \eqref{eq:location_scale}, assume that $\sigma^*(X) = \exp(\gamma^{\top}X)$ for some $\gamma \in \mathbb{R}^d$. Then, LogS and SCRPS proper scoring rules admit an invariant prediction with respect to the class of interventional distributions $P$, generated as in \eqref{eq:perturbation}, for which
\[
    \mathbb{E}_P(X) - \mathbb{E}_{P^o}(X) \perp \gamma.
\]
\end{lemma}
For heteroscedastic linear models, the exponential scale $\sigma^\ast(X)$ is a popular assumed parametric form for the variance term that, to the best of our knowledge, was first considered by \cite{cook1983diagnostics}.  For general location-scale models with linear scale parametrization and exponential link, interventions under which the expectation of $X$ remains constant or is shifted in a direction orthogonal to $\gamma$ do not affect the risk with respect to LogS and SCRPS. Clearly, $\pi^*_{y|\cdot}$ is not the unique $(S, \mathcal{P})$-invariant probabilistic prediction under the distributions in Lemma \ref{lem:exponential_scale}. For example, if $\pi'_{y|x} = \mathcal{L}\{\mu^*(x) + \sigma^*(x) \eta\}$ and $\eta$ follows an arbitrary distribution $Q'\neq Q^o$, \revision{then the risk is also invariant. However, since $\mathbb{E}[\log\{\sigma^*(x)\}]$ is the same in $S(\pi'_{y|X})$ and $S(\pi^*_{y|x})$, and $\mathrm{LogS}(Q', \varepsilon_Y) > \mathrm{LogS}(Q^o, \varepsilon_Y)$, the prediction $\pi^*_{y|\cdot}$ achieves a smaller risk than location-scale families with a misspecified baseline distribution.}  
In general it is not obvious whether there exist other predictions with misspecified location or scale functions that achieve a smaller, invariant risk than $\pi^*_{y|\cdot}$, since our setting generally does not allow do-interventions and Proposition \ref{prop:invariance_probabilistic} does not apply. The following example shows that in the Gaussian heteroscedastic linear regression model with LogS, one can indeed identify $\pi^*_{y|\cdot}$ as the unique prediction method with a minimal invariant risk.

\begin{example} \label{ex:identifiability}
Assume that $(\varepsilon_Y, \varepsilon_X^{\top})^\top$ are Gaussian with mean $0$ and covariance matrix $\Sigma \in \mathbb{R}^{(d+1)\times (d+1)}$ with $\Sigma_{11} = 1$. Let $\mu^*(x) = \beta^{\top}x$ and $\sigma^*(X) = \exp(\gamma^{\top}x)$, and let
\[
    \pi_{y|x} = \mathcal{N}\{b^{\top}x, \exp(2g^{\top}x)\}
\]
for some $b, g \in \mathbb{R}^d$. We consider finitely many interventions on the covariance structure of $X$ to formulate identifiability conditions; more precisely, we assume
\[
    X^e = \Gamma^e \varepsilon_X, \quad Y^e \leftarrow \beta^{\top}X^e + \exp(\gamma^{\top}X^e)\varepsilon_Y,
\]
where $e = 1, \dots, E$ are the observations under $E$ environments as in model \eqref{eq:environments},
with interventions that are multiplication of $\varepsilon_X$ with some invertible matrix $\Gamma^e$. Then,
\begin{align*}
    2\, \mathrm{LogS}(P_{y\mid X^e}, Y^e) & = \left[\varepsilon_Y\exp\{(\gamma-g)^{\top}\Gamma^e\varepsilon_X)\} + (\beta-b)^{\top}\Gamma^e\varepsilon_X\exp(-g^{\top}\Gamma^e\varepsilon_X)\right]^2 \\
    & + g^{\top}\Gamma^e\varepsilon_X + c,
\end{align*}
where $c$ is a constant not depending on the parameters.  
Let $\Sigma_X = (\Sigma_{ij})_{i,j=2}^{d+1}$ be the covariance matrix of $\varepsilon_X$ and $\Sigma_{YX} = (\Sigma_{i1})_{i=2}^{d+1}$ the covariance between $\varepsilon_X$ and $\varepsilon_Y$.
Since the mean of $\varepsilon_X$ is zero, we have $\mathbb{E}(g^{\top}\Gamma^e\varepsilon_X) = 0$, and the expected score equals
\begin{align*}
    & 2 \,\mathbb{E}(\mathrm{LogS}(P_{y\mid X^e}, Y^e)) \\
    & =  \left[(\beta - b)^{\top}\Gamma^e \Sigma_X (\Gamma^e)^{\top}(\beta - b) + 4\{(\beta - b)^{\top}\Gamma^e \Sigma_X (\Gamma^e)^{\top}g\}^2\right]\exp(2g^{\top}\Gamma^e \Sigma_X (\Gamma^e)^{\top}g) \\
	& \quad + \exp\{2(\gamma - g)^{\top}\Gamma^e \Sigma_X (\Gamma^e)^{\top}(\gamma - g)\} \revision{[1+\{(\gamma - g)^{\top} \Gamma^{e} \Sigma_{YX}\}^{2}]} \\
	& \quad + 2(\beta - b)^{\top}\Gamma^e \Sigma_{YX} \exp\{(\gamma - 2g)^{\top}\Gamma^e \Sigma_X (\Gamma^e)^{\top} (\gamma - 2g)/2\} + c
\end{align*}
From the above calculations, we see that one of the following are sufficient conditions for $\beta$ and $\gamma$ to be uniquely identified as the parameters attaining the minimal invariant risk.

\begin{condition}\label{it:no_confounding}
If there is no confounding, meaning that $\Sigma_{YX} = 0$, then the minimal risk is always achieved uniquely by $b = \beta$ and $g = \gamma$. This follows from the formula above, where the third term vanishes, or directly from strict propriety of the logarithmic score, and invariance holds because $\Gamma^eX$ has mean zero for all $e$.
\end{condition}

\begin{condition}\label{it:covariance_condition}
If $b = \beta$, or if there is no location parameter ($\beta = b = 0$), then only $g = \gamma$ achieves a minimal constant risk if there are two environments for which $\Gamma^e\Sigma_X(\Gamma^e)^{\top} \prec \Gamma^{e'}\Sigma_X(\Gamma_{e'})^{\top}$, i.e., $\Gamma^{e'}\Sigma_X(\Gamma^{e'})^{\top} - \Gamma^{e}\Sigma_X(\Gamma^e)^{\top}$ is strictly positive definite.
\end{condition}

\begin{condition}\label{it:different_environments}
Assume that $E \geq d + 1$ and there exist $d$ environments $e_1, \dots, e_d$ such that $\Gamma^{e_j}\Sigma_{YX}$, $j = 1, \dots, d$, are linearly independent, and an environment $e_{d+1}$ such that $\Gamma^{e_{d+1}}\Sigma_{YX} = \sum_{j = 1}^d \alpha_j \Gamma^{e_{j}}\Sigma_{YX}$ for some $\alpha_j \leq 0$, $j = 1, \dots, d$. Then the risk is constant and minimal only if $b = \beta$ and $g = \gamma$. 
\end{condition}

\end{example}

\revision{Conditions \ref{it:no_confounding}, \ref{it:covariance_condition} and \ref{it:different_environments} are different scenarios under which the do-interventional distributions $\pi^*_{y|\cdot}$, or the corresponding parameters $\beta$ and $\gamma$, are identified through invariance and minimality of the risks over finitely many environments. Condition \ref{it:no_confounding} achieves this by assuming that there is no confounding between $X^e$ and $Y^e$. A similar condition appears in \citet[Proposition 3.1]{Christiansen2021}, who obtain invariance and robustness of the causal function if there is at least one intervention that removes confounding, like do-interventions. The condition is rather strong, since, as we have seen in Theorem \ref{thm:score_not_exist}, one usually cannot expect invariance of risks under such strong interventions. Condition \ref{it:covariance_condition} ensures identifiability if the covariance matrices of $X^e$ are ordered in the Lowener order for two environments. A similar modelling assumption is made by \citet[Section 2.1]{Shen2023} to derive a robust estimator under interventions. And finally, our Condition \ref{it:different_environments} ensures identifiability if there are sufficiently many environments between which the confounding}, i.e., the covariance between $X^e$ and $\varepsilon_Y$, varies strongly enough. It is a natural identifiability condition, since it should be difficult to separate the effect of $X^e$ on $Y^e$ from the confounding if there is a similar confounding in all environments.}

\revision{\subsection{Prediction intervals}
So far, we have analyzed the invariance and robustness of predictions in terms of their forecast error. Since probabilistic forecasts directly allow  constructing prediction intervals by taking quantiles of the forecast distribution, a further question of interest is whether robustness also translates to coverage guarantees of prediction intervals under interventions. The following proposition shows that the do-interventional distributions $\pi^*_{y|\cdot}$ indeed have such guarantees.
\begin{proposition}\label{prop:coverage}
    Under the model \eqref{eq:location_scale}, let $q_{\alpha/2}$ and $q_{1-\alpha/2}$ denote the $\alpha/2$ and $1-\alpha/2$ quantiles of $\varepsilon_{Y}$ respectively and define for any $x \in \mathbb{R}^d$
    \begin{align*}
        \ell^{\ast}_{x}(\alpha) &= \mu^{\ast}(x) + \sigma^{\ast}(x) q_{\alpha/2} \\
        u^{\ast}_{x}(\alpha) &= \mu^{\ast}(x) + \sigma^{\ast}(x) q_{1-\alpha/2}.
    \end{align*}
    Then,
    \[
    \mathbb{P}\{Y^{\mathrm{int}} \in [\ell^*_{X^{\mathrm{int}}}(\alpha), u^*_{X^{\mathrm{int}}}(\alpha)]\} \geq 1-\alpha
\]
under all interventions of the form \eqref{eq:perturbation}, with equality if the distribution of $\varepsilon_Y$ is continuous.
\end{proposition}
We emphasize that the coverage guarantee in Proposition \ref{prop:coverage} is computed over the joint distribution of $(X^{\mathrm{int}}, Y^{\mathrm{int}})$: since these variables are from an interventional distribution, this is a different guarantee than a marginal guarantee over the training data from $(X,Y)$ which appears in the conformal inference literature \citep{Tibshirani2019, Cauchois2024, Ai2024}.} 

\section{Invariant probabilistic prediction}\label{sec:ipp}
The derivations and examples in the previous sections suggest that, in order to find an invariant and thus robust probabilistic prediction, one should seek for a $\pi_{y|\cdot}$ that achieves a minimal, constant risk across different interventional distributions. 
In this section, we consider the model for multiple environments given in \eqref{eq:environments}, with each environment $e=1,\dots,E$ corresponding to one distinct intervention. 

Assume that the candidate predictions $\pi_{y|\cdot}$ are parametrized by some $\theta \in \Theta \subset \mathbb{R}^d$, for which we write $\pi_{y|\cdot, \theta}$. 
For example, in the exponential location-scale model from Lemma \ref{lem:exponential_scale}, we have $\theta = (\beta^\top, \gamma^\top)^\top$.  
Let $S$ be a strictly proper scoring rule and define the empirical risk and its expectation in environment $e$ by
\[
    \hat{\mathcal{R}}^e(\theta) = \frac{1}{n_e}\sum_{i=1}^{n_e} S(\pi_{y|X^e_i,\theta}, Y^e_i), \quad \mathcal{R}^e(\theta) = \mathbb{E}(\hat{\mathcal{R}}^e) = \mathcal{R}( \pi_{y|\cdot, \theta}, P^e, S),
\]
where $(X^e_i, Y^e_i)$, $i = 1, \dots, n_e$, are independent and identically distributed observations in environment $e$. It is always assumed that the expectations above are finite. Now, we can define our proposed estimator, invariant probabilistic prediction.

\begin{definition}[Invariant probabilistic prediction (IPP)] \label{def:ipp}
Let $S = S(P,Y)$ be a strictly proper scoring rule, $D\colon\mathbb{R}^E \rightarrow [0,\infty)$ be continuous a function such that $D(v) = 0$ if and only if $v_1 = \dots = v_k$, $\lambda \geq 0$ be a tuning parameter, and $\{\omega^e\}_{e=1}^{E}$ be convex weights.

The invariant probabilistic prediction is defined as $\pi_{y|\cdot, \hat{\theta}(\lambda)}$, where $\hat{\theta}(\lambda)$ minimizes 
\begin{equation} \label{eq:target}
    \hat{R}_{\lambda}(\theta) = \sum_{e=1}^E\omega^e\hat{\mathcal{R}}^e(\theta) + \lambda D\{\hat{\mathcal{R}}^1(\theta), \dots, \hat{\mathcal{R}}^E(\theta)\},
\end{equation}
if a minimizer exists.
\end{definition}

The weights can be for example uniform weights or proportional to the sample size per environment. 
Our proposed IPP estimator turns out to be a generalization of the V-REx estimator of \cite{Krueger2021}, who consider the special case where the convex weights are uniform and the regularization function is the variance,
\begin{align}\label{eq:variance}
    D(v) = \sum_{1 \leq i < j \leq E}(v_i - v_j)^2 / E^2.
\end{align}
However, IPP is derived in a more formal and general manner by characterizing the invariance of probabilistic prediction, and its theoretical properties are well demonstrated. 
In particular, the following theorem shows that, under mild regularity conditions, IPP yields a consistent estimator of the underlying parameter and an invariant prediction.

\begin{theorem} \label{thm:consistency}
Consider the model given in \eqref{eq:environments}.
Suppose \vspace{-0.1in}
\begin{enumerate}
    \item there exists a unique $\theta^\ast \in \Theta$ such that $\mathcal{R}^1(\theta^*) = \dots = \mathcal{R}^E(\theta^*)$ and $\mathcal{R}^1(\theta^*)$ is minimal among the parameters achieving an invariant risk; \label{it:identifiability}
    \item the number of observations in each environment satisfies $n := \min(n_1, \dots, n_E) \to \infty$; \label{it:sample_size}
    \item the parameter space $\Theta$ is compact; \label{it:compactness}
    \item the function $S(\pi_{y|x,\theta},y)$ is continuous in $\theta$ for all $(x,y)$; \label{it:continuity_score}
    \item there exists $0 < B < \infty$ such that $\mathbb{E}\{\sup_{\theta\in\Theta}S(P_{y|X^e,\theta}, Y^e)^2\} \leq B$, $e = 1, \dots, E$; \label{it:moment}
    \item \label{it:reg} the sequence $\lambda_{n} \to \infty$ and the regularization function $D$ satisfy 
    \begin{align*}
        \sup_{\theta\in\Theta}\lambda_n|D\{\hat{\mathcal{R}}^1(\theta), \dots, \hat{\mathcal{R}}^E(\theta)\} - D\{\mathcal{R}^1(\theta), \dots, \mathcal{R}^E(\theta)\}| = o_P(1),
    \end{align*}
\end{enumerate}\vspace{-0.1in}
then $\hat{\theta}(\lambda_{n})$ is consistent for $\theta^{\ast}$.  If, in addition, \vspace{-0.1in}
\begin{enumerate}
    \setcounter{enumi}{6}
    \item the family $\{\pi_{y|x, \theta}\}$ is continuous in $\theta$ for a fixed $x$, \label{it:p_cont}
\end{enumerate}\vspace{-0.1in}
then $\pi_{y|x, \hat{\theta}(\lambda_n)}$ is consistent for $\pi_{y|x, \theta^{\ast}}$.  
\end{theorem}

The first assumption ensures that the invariant population parameter $\theta^{\ast}$ is identified and, effectively, imposes a restriction on the class of interventions.  As demonstrated in Example \ref{ex:identifiability}, this assumption is satisfied for the Gaussian heteroscedastic linear regression model with $\mathrm{LogS}$, while such a specific model is not required in our consistency result.  Assumption \ref{it:sample_size} requires the number of observations in each environment to diverge, to ensure sufficient information about all of the environments.  Assumptions \ref{it:compactness}, \ref{it:continuity_score}, \ref{it:moment}, and \ref{it:p_cont} are standard conditions for consistency. For instance, Assumptions \ref{it:continuity_score} and \ref{it:moment} are satisfied in Example \ref{ex:identifiability} if the parameter space is compact. 

The remaining Assumption \ref{it:reg} concerns the penalty.  In particular, the choice of $\lambda_{n}$ depends on the regularization function $D$.   \revision{In the simple case where we have a single environment and use LogS, our assumptions are similar to those required for consistency of maximum likelihood; for example, see \cite{wald1949}.}
The following corollary gives an explicit characterization of this assumption in the setting where $D$ is the variance function in \eqref{eq:variance}.

\begin{corollary} \label{cor:penaltypar}
    Consider the model given in \eqref{eq:model}.  Suppose Assumptions \ref{it:identifiability} -- \ref{it:moment} and \ref{it:p_cont} hold.  If $D$ is the variance function given in \eqref{eq:variance} and $\lambda_{n} \to \infty$ with $\lambda_n = O\{n^{1/2}\log(n)^{-1/2}\}$, then Assumption \ref{it:reg} is satisfied.  Moreover, the conclusions of Theorem \ref{thm:consistency} hold.
\end{corollary}

Theorem \ref{thm:consistency} only gives an upper bound on the rate at which $\lambda$ may diverge, but it does not give guidance of how to choose $\lambda$ in applications. Our strategy, which is inspired by \citet{Jakobsen2021} and yields good results in our experiments, is to select a minimal $\lambda$ for which there are no significant differences between the environment risks. More precisely, let $\hat{T}(\theta)$ be a test statistic depending on $\theta$ and $(X_i^e, Y_i^e)$, $i = 1, \dots, n_e$, $e = 1, \dots, E$, such that higher values of $\hat{T}(\theta)$ indicate a violation of the hypothesis $\mathcal{R}^1(\theta) = \dots = \mathcal{R}^E(\theta)$. Assume that the (asymptotic) distribution function $F$ of $\hat{T}(\theta)$ in the case that $\mathcal{R}^1(\theta) = \dots = \mathcal{R}^E(\theta)$ is known. Then for $\alpha \in (0,1)$, the set $C(\alpha) = \{\theta \in \Theta\colon F\{\hat{T}(\theta)\} \leq 1 - \alpha\}$ is an (asymptotic) confidence set for the parameter $\theta^*$. We then choose $\lambda$ as
\[
    \hat{\lambda} = \inf\{\lambda \geq 0\colon \hat{\theta}(\lambda) \in C(\alpha)\},
\]
i.e., we increase the penalty $\lambda$ until the sample estimator $\hat{\theta}_{\lambda}$ lies in the confidence set $C(\alpha)$. Hence the choice of $\lambda$ is reduced to finding the minimal penalty parameter for which the null hypothesis of equal risks is not rejected, for a small confidence level $\alpha$. Notice that there is no guarantee that the p-value $F(\hat{T}(\hat{\theta}_{\lambda}))$ is increasing in $\lambda$, but in practice one observes this in most cases due to the following simple result. 

\begin{lemma} \label{lem:penalty_decreasing}
The function $\lambda \mapsto D[\hat{\mathcal{R}}^1\{\theta(\lambda)\}, \dots, \hat{\mathcal{R}}^E\{\theta(\lambda)\}]$ is non-increasing in $\lambda$, and $\lambda \mapsto \sum_{e=1}^E\hat{\mathcal{R}}^e\{\theta(\lambda)\}$ is non-decreasing in $\lambda$. 
\end{lemma}

In our empirical applications, we use the generalization of Welch's two sample t-test with unequal variances implemented in the function {\tt oneway.test} in \texttt{R}. The weights in Definition \ref{def:ipp} are taken as $\omega^e = 1/E$, $e = 1, \dots, E$, and for the function $D$, we choose the variance \eqref{eq:variance} for computational simplicity. 

\revision{
For the proper scoring rules, we focus on the logarithmic score and the SCRPS, since these are the scores for which we can formulate plausible statistical models under which invariance holds; see Section \ref{sec:restricted_interventions}. \citet{Bolin2023} show that these scores are locally scale invariant in the sense that the prediction error, relative to an ideal forecast, is independent of the scale of the observations, which may be a desirable property if the scale of the observations differs between environments. This is not the case for other scores, e.g., CRPS and Hyv\"arinen score, and we refer to \citet{Bolin2023} for the details.
In general, the choice of a suitable proper scoring rule is a non-trivial problem and depends on the application at hand, as discussed by \citet[Section 2.1]{Carvalho2016}. For instance, the logarithmic score is known to be more strongly influenced by single outcomes in low predictive density regions, whereas such observations are less influential in the CRPS, quadratic, and spherical score \citep{Machete2013}. Apart from the theoretical considerations above, practical aspects also often guide the choice of the scoring rule. The CRPS and SCRPS are suitable for models that allow sampling observations but that have no simple closed form densities; if a density is available, the logarithmic score is often easiest to compute. In Section \ref{sec:single_cell} and in Appendix \ref{sec:sim2}, we also present results with the CRPS as scoring rule. Generally, our empirical results suggest that invariant predictions with respect to a certain proper scoring rules often also exhibit relatively small dispersion with respect to other scoring rules; thus, in practice, the choice of the scoring rule has a less strong effect than the choice of the prediction method.
}

\section{Empirical results}\label{sec:empirical}

\subsection{Simulations}\label{sec:simulations}
We illustrate IPP in the setting of Example \ref{ex:identifiability}. For $d = 5$, we generate $(\varepsilon_Y, \varepsilon_X) \sim \mathcal{N}_{d+1}(0, \Sigma)$, where
\begin{align*}
    \Sigma_{ii} = 1, i = 1, \dots, d + 1, \
    \Sigma_{12} = 0.8, \ \Sigma_{13} = -0.4, \ \Sigma_{14} = 0.3, \ \Sigma_{15} = -0.2, \ \Sigma_{16} = 0.1,
\end{align*}
and $\Sigma_{ij} = 0$ for all other $i, j$.
That is, $\varepsilon_X$ has independent standard normal entries, but there is confounding, due to the non-zero correlations with $\varepsilon_Y$. In environments $e = 1, \dots, d$, we define
\begin{align*}
    & X^e = \Gamma^e \varepsilon_X, \, (\Gamma^e)_{ij} = 1(i=j) + U_{ij}, \, U_{ij} \sim \mathrm{Unif}(-0.1, 0.1), \\
    & \beta = (\beta_1, \dots, \beta_{d}), \, \beta_j \sim \mathrm{Unif}(0,3), \quad \gamma = (\gamma_1, \dots, \gamma_{d}), \, \gamma_j \sim \mathrm{Unif}(0,0.5), \ j = 1, \dots, d, \\
    & Y^e = \sum_{j=1}^d X_j^e\beta_j + \exp\Big(\sum_{j=1}^d X_j^e \gamma_j \Big)\varepsilon_Y,
\end{align*}
where all the random parameters are chosen independently. With this construction, the covariance of $X$ is slightly different in each environment but close to the identity matrix. For environment $d+1$ the data is generated in the same way, but $\Gamma^{d+1} = \sum_{e = 1}^{d} \alpha^e \Gamma^e$
for random $\alpha^e \leq 0$ summing to $-1$. Condition \ref{it:different_environments} in Example \ref{ex:identifiability} is satisfied with probability one in this simulation study, \revision{and the interventions simultaneously affect the marginal covariate distribution and the conditional distribution of the response variable given the covariates.} We consider $\lambda = 0, 0.5, 1, \dots, 15$ and sample sizes of $n^e = 100, 150, 200, 250, 500, 1000$ in each environment. Estimation is done with $S = \mathrm{LogS}$ and $S = \revision{\mathrm{SCRPS}}$, and the figures for the latter are given in Appendix \ref{sec:add_figures}.

We implemented our methods in \texttt{R} and \texttt{Python}, and provide code and replication material for all our empirical results on \url{https://github.com/AlexanderHenzi/IPP}. Details about the computation of IPP are given in Appendix \ref{sec:experimental_details}.

\begin{figure}[t]
    \centering
    \includegraphics[width=0.9\textwidth]{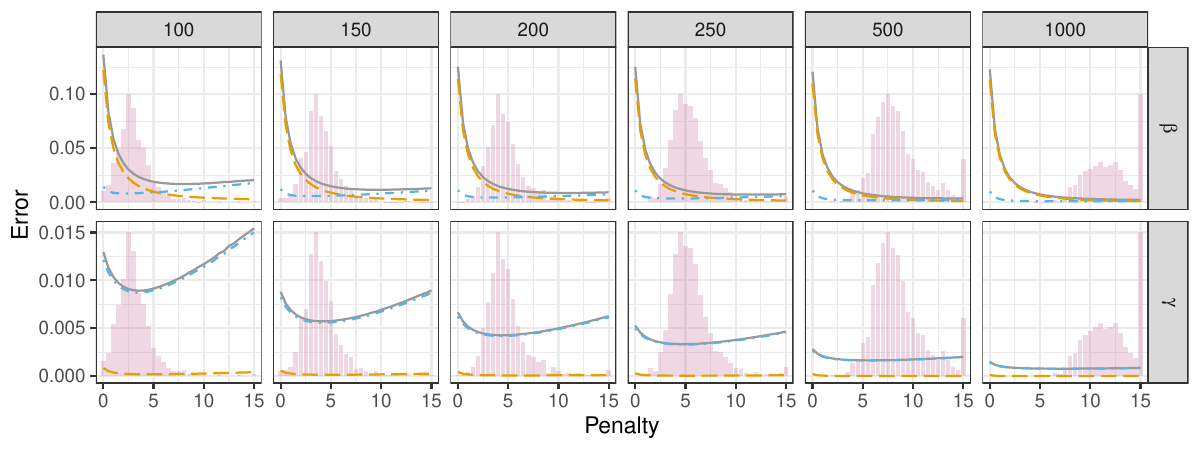}\vspace{-0.15in}
    \caption{Simulation study: Estimation error for $\beta$ (top) and $\gamma$ (bottom) as a function of penalty parameter, separated into squared bias (blue, dot-dashed), variance (orange, dashed), and total error (gray, solid), for different $n^e$. The sample sizes are given on top of each panel. The bars show height-adjusted frequencies with which $\lambda$ is chosen by the rule described in Section \ref{sec:ipp} with $\alpha = 0.05$; $\lambda$ is set to the maximal value $15$ for this experiment if the hypothesis of equal risk is rejected for all $\lambda$. \label{fig:parameter_error}}
\end{figure}

Figure \ref{fig:parameter_error} shows the mean parameter estimation errors $\mathbb{E}\{\|\hat{\beta}_{\lambda} - \beta\|^2\}$ and $\mathbb{E}\{\|\hat{\gamma}_{\lambda} - \gamma\|^2\}$, and their decomposition into bias and variance. For the location parameter, regression without penalization is biased, and as $\lambda$ increases, the bias decreases. For the scale parameter the bias is negligible but the variance decreases for moderate $\lambda$. Choosing $\lambda$ too large deteriorates the estimator for both parameters. Our heuristic rule for choosing the $\lambda$ with $\alpha = 0.05$ often selects a penalty parameter that gives small estimation errors.  Levels of $\alpha = 0.01$ or $\alpha = 0.1$ yield slightly smaller or greater $\lambda$, respectively, but qualitatively the same results; see the figures in Appendix \ref{sec:add_figures}.

\begin{figure}[t]
    \centering
    \includegraphics[width=0.9\textwidth]{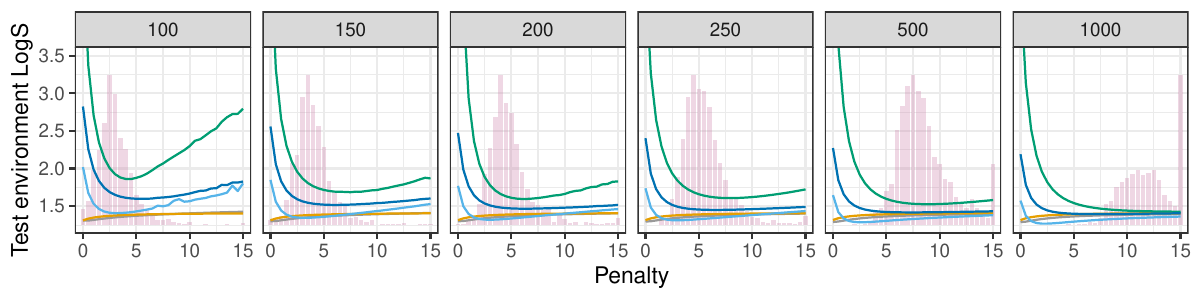}\vspace{-0.1in}
    \caption{Simuation study: Mean logarithmic score on new test environment arising from 
    the high and low variance interventions (light blue, orange), mean shift (dark blue), intervention on correlation (green), and without interventions compared to the observational distribution (gray). \revision{The histograms are height-adjusted frequencies of the chosen penalty parameter, as in Figure \ref{fig:parameter_error}.}  \label{fig:interventions}}
\end{figure}

In Figure \ref{fig:interventions} we depict the mean logarithmic score when predicting on test environments. If the data in the test environment has the same distribution as the pooled training data, then distributional regression without penalization is clearly the best choice. We then tested variance interventions, where $X^e = c^e \varepsilon_X$ for $c^e = 1/3, \, 3/2$, labelled as low and high variance interventions, respectively. Furthermore, we applied an intervention on the correlation structure, where
\[
    X^e = \Gamma^e \varepsilon_X, \ \Gamma_{ij}^e \sim 1(i = j) + \mathrm{Unif}(-0.75, 0.75), \ i,j = 1, \dots, d,
\]
and a mean shift intervention orthogonal to $\gamma$,
\[
    X^e = \varepsilon_X + \delta^e - \gamma \{(\delta^e)^{\top}\gamma\} / \|\gamma\|^2, \ \delta^e_j \sim \mathrm{Unif}(-5, 5), \ j = 0, \dots, d.
\]
While IPP is not guaranteed to have a minimal error under all interventions, we observe that it does minimize the maximum error among the classes of interventions considered, in comparison to distributional regression without penalization. Also, it can be seen that the error under interventions for large $\lambda$ is almost constant across different interventions, at least for sufficiently large sample sizes, which is not the case for distributional regression.

\revision{Further simulations, where we compare IPP against other methods and also use the CRPS for training and evaluation, can be found in Appendix \ref{sec:add_figures}.}

\subsection{Application on single-cell data} \label{sec:single_cell}
So far we have demonstrated that IPP achieves, under ideal conditions, an invariant risk under certain classes of interventions on the covariates. In practice, one would hope that even in a misspecified setting or when interventions do not only act on the covariates, equalizing training risks still has a positive effect on the out-of-distribution performance in new environments. We investigate this in an application on single-cell data, similar to \citet[Section 6]{Shen2023}. Our data consists of measurements of the expression of 10 genes, one of which is chosen as the response variable and the others as covariates. The training data comprises 9 environments, where each of the  covariate genes is intervened on with CRISPR pertubations, and an additional observational environment without interventions. The number of observations range from $101$ to $552$ for the interventional environments with the perturbed genes, and equals $11485$ for the observational environment. In addition to the training data, observations from more than $400$ environments with interventions on other genes are available. We evaluate predictions on the 50 environments in which the distribution of the 10 observed genes has the largest energy distance \citep{Szekely2004} from their distribution in the pooled training data.

\begin{figure}
    \centering
    \includegraphics[width=0.9\textwidth]{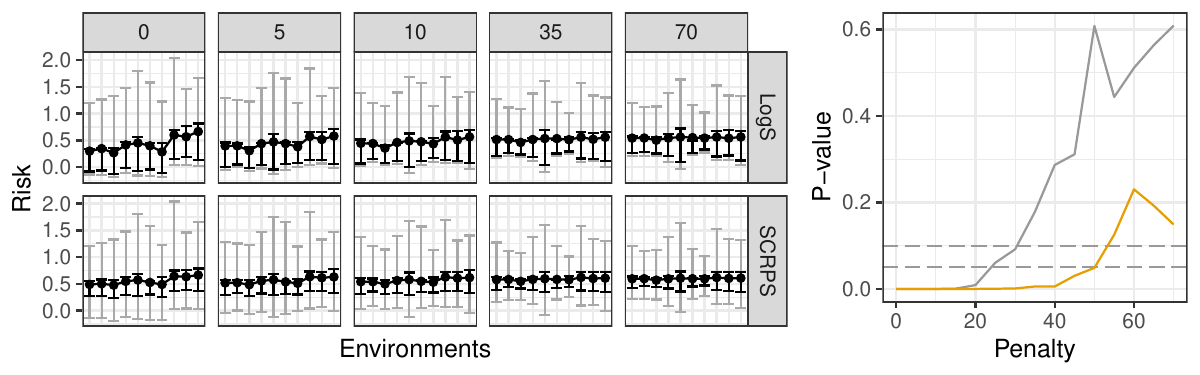}\vspace{-0.1in}
    \caption{Single cell data: Training risks of IPP over different environments (left panel)\revision{; the dot is the mean risk, the inner whiskers the $0.25$- and $0.75$-quantile of the scores, and the outer whiskers the $0.05$- and $0.95$-quantile}. The first environment is the observational environment, and the panel columns are different penalty parameter values. The right panel shows the p-value of the {\tt oneway.test} for equal training risks as a function of the penalty, in gray for LogS and orange for SCRPS Horizontal lines indicate the levels of $0.05$ and $0.1$. \label{fig:stability_pvals}}
\end{figure}

We first apply IPP with the heteroscedastic Gaussian linear model,
\[
    \pi_{y|x} = \mathcal{N}\{\beta_0 + \beta^{\top}x,\, \exp(2\gamma_0 + 2\gamma^{\top}x)\},
\]
where $x$ is the vector of expressions of the $9$ covariate genes, and perform estimation with LogS and SCRPS as loss functions. Figure \ref{fig:stability_pvals} illustrates the training risks of above versions of IPP as a function of the penalty parameter, and the corresponding p-values for testing equality of training risks. For the LogS, a significance level of $\alpha$ between $0.05$ and $0.1$ would select a penalty parameter between $20$ and $30$. For the SCRPS a value of about $50$ is required to achieve non-significant difference between training risks.

As an extension, we apply IPP with the model \eqref{eq:model} in full generality. Specifically, we adopt a model $Y = f_{\theta}(X,\varepsilon)$ with a nonlinear function $f_{\theta}(X, \varepsilon)$ parametrized by neural networks, taking as arguments the covariates $X$ and multivariate Gaussian $\varepsilon$ representing the noise. A probabilistic prediction $\pi_{y|x}$ is obtained by resampling the noise $\varepsilon$ while fixing $X=x$.  Such general parametrizations of probabilistic predictions have been used in the literature~\citep{shen2023engression}, and due to the flexibility of the function class for $f_{\theta}$, the choice of the distribution of $\varepsilon$, here specified as Gaussian, usually has no impact on the outcome distribution.
In this general setting, the conditional distribution does not have a closed-form density so we cannot evaluate the LogS explicitly, but the generative nature of this model enables estimation of the SCRPS by sampling.
Thus, we apply IPP with the SCRPS as the loss function, and refer to this as a neural network implementation (NN) of IPP with SCRPS. \revision{Implementation details for all methods in this section are in Appendix \ref{sec:experimental_details}.}

As a competitor for probabilistic prediction, we apply distributional anchor regression \citep{Kook2022} with $F_{y|x}(z) = \Phi\{h(z) - \beta^{\top}x\}$ for a strictly increasing function $h$ parametrized with Bernstein polynomials and estimated jointly with $\beta$. Due to optimization problems with large data sets, we had to exclude the observational environment for this method, and to ensure a fair comparison, Appendix \ref{sec:add_figures} includes a figure of the prediction error when IPP is also applied without the observational environment, where the advantage over distributional anchor regression remains similar. \revision{Furthermore, we generate conformal prediction intervals with the methods by \citet{Cauchois2024} and \citet[Algorithm 2]{Ai2024}, applied on only the observational environment, abbreviated as ``Conformal (obs)'' and ``Conformal (weighted, obs)'' in the figures. In Appendix \ref{sec:add_figures}, we also test the effect of pooling all training environments for conformal prediction. Unlike all other methods, \citet{Ai2024} require samples of the test data covariates during training.}

In addition to the distributional methods, we reduce the probabilistic predictions to their conditional means and compare with respect to the mean squared error (MSE) against methods aimed for mean prediction, such as anchor regression \citep{Rothenhausler2021} and DRIG \citep{Shen2023}. The setting for distributional anchor regression, anchor regression, and DRIG does allow interventions on the response variable, and we investigate the influence of including the corresponding environment in the training data Appendix \ref{sec:add_figures}. Finally, as another baseline method for point prediction, we test V-REx \citep{Krueger2021} with $y = f_{\theta}(x)$ parametrized by neural networks with the same architecture as our neural network implementation of IPP. In this method, the loss function is of the same form as in IPP except replacing SCRPS with the squared error loss, i.e.,  $\mathcal{R}^e(\theta)=\bbE[\{Y^e-f_{\theta}(X^e)\}^2]$. 

\begin{figure}
    \centering
    \includegraphics[height=0.85\textheight]{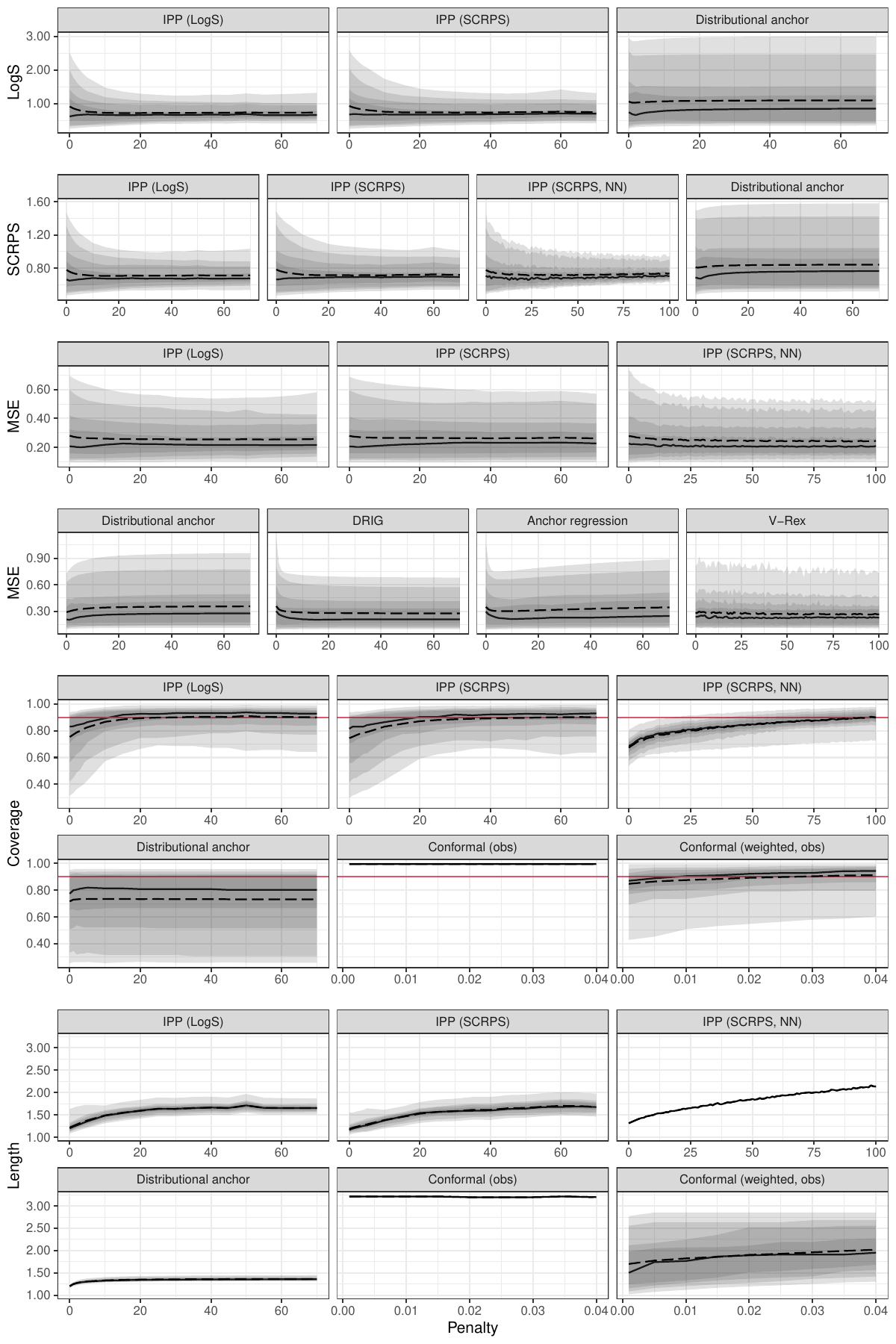}\vspace{-0.1in}
    \caption{Single cell data: \revision{Evaluation metrics} on the $50$ test environments. The solid and dashed line show the median and mean environment prediction error, and the shaded bands are delimited by the $j/10$- and $(10-j)/10$-quantiles, $j = 0, \dots, 4$ of the mean \revision{evaluation metrics} over the $50$ environments, respectively. Note that the conformal methods are not directly comparable: ``Conformal (obs)" and ``Conformal (weighted, obs)" are based on the observational training environment while the latter uses additional knowledge of the covariates from the test data.
    \label{fig:single_cell_test}
    }
\end{figure}

Fig.~\ref{fig:single_cell_test} depicts \revision{measures of prediciton quality on the test data, namely, LogS and SCRPS as well as $90\%$-prediction interval coverage and length} for distributional methods, and MSE for all methods to evaluate the mean prediction. IPP always stands as the most competitive method regardless of the loss function and model class used, as it consistently achieves the smallest prediction errors, especially the worst-case error, and also the smallest variation in the prediction errors across all test environments. It exhibits similar advantages even in terms of point prediction for the mean over the methods designed for mean prediction. Interestingly, when heteroscedastic Gaussian linear models are used, estimating IPP with the logarithmic score as the loss function outperforms the estimation with SCRPS, even when evaluated with respect to the latter score, suggesting that a more classical approach with log-likelihood as loss function fits the model better. \revision{For suitable penalization the coverage of the IPP $90\%$-prediction intervals is close to the nominal level, on average, and the worst case coverage over the test environments is better than that of distributional anchor regression and conformal prediction, while having similar average length as those by \citet{Ai2024}. The intervals by \citet{Cauchois2024} are overly wide and conservative, which is observed empirically by \citet{Ai2024}. We observe that the length of the prediction intervals by \citet{Ai2024} varies the strongest between test environments, whereas distributional anchor regression and the neural network variant of IPP generate more homogeneous lengths; the parametric variant of IPP has slightly more variability, but far less than weighted conformal prediction. The conformal prediciton intervals by \citet{Cauchois2024} all have the same length by construction. Further comparisons, also in terms of CRPS, are in Appendix \ref{sec:add_figures}}

A suitable penalization strength generally improves all methods. The penalty parameters selected for IPP in the heteroscadastic Gaussian linear version, based on the p-values in Fig.~\ref{fig:stability_pvals}, yield a stable and small error both under LogS and SCRPS, showing the usefulness of our scheme for selecting the penalty. The effect of penalty tuning is the weakest for distributional anchor regression, where a small but positive penalty parameter improves the test errors only slightly. 

With the general model class using neural network parametrization, IPP exhibits slightly better performance compared to the heteroscedastic Gaussian linear model, especially with a larger penalty parameter. The performance across different test environments also tends to be less variable, indicated by the smaller spreads. Notably, this improvement does not come only from the flexibility of the model class. In fact, the V-REx method, which is also based on neural networks, is inferior to the NN variant of IPP in terms of both the average risk and variability, in spite of the fact that it is optimized to have small and constant risks in terms of MSE on training data. This highlights that even if one is interested in point prediction, it may be beneficial to apply distributional methods and then deduce any summary statistic such as the mean or the median.

\section{Discussion}
We considered and analyzed invariance for probabilistic predictions under structural equation models, aka a causality-inspired framework \citep{peters2016causal,Rojas2018,buhlmann2020invariance}. In contrast to point estimation, which has been studied in a lot of recent works, the rigorous formulation for probabilistic predictions has not been worked in detail. We described the fundamental difficulties and brought up a constructive algorithm, called Invariant Probabilistic Prediction (IPP), which is shown to have desirable theoretical properties, as well as exhibiting good and robust performance on simulated and real data. We believe that our fundamental formulation would lead to some avenues of future research in various directions, including more detailed robustness against a finite amount of distribution shifts, instead of full invariance, or rigorous analysis of invariance and identifiability in more complex nonlinear models.

\section*{Acknowledgement}
We are grateful to Lucas Kook for advice on the implementation of distributional anchor regression, \revision{and to the associate editor and a referee for helpful comments.}  AH and PB are supported in part by European Research Council (ERC) under European Union's Horizon 2020 research and innovation programme, grant agreement No. 786461.  XS is supported in part by the ETH AI Center.  ML is supported in part by NSF Grant DMS-2203012.  

\setlength{\bibsep}{0pt plus 0.3ex}
\bibliography{references}
\bibliographystyle{plainnat}

\appendix
\section{Proofs} \label{sec:proofs}

\subsection{Proof of Proposition 1}
\begin{proof}
The first equality holds because the distribution of $L\{\mu^*(X), Y\}$ only depends on $\varepsilon_Y$, which is invariant under interventions on $X$. For the second equality, notice that for any $\mu$,
\[
    \sup_{P \in \mathcal{P}} \mathbb{E}_P\{L(\mu(X), Y)\} \geq \mathbb{E}_{\mathrm{do}(X = x)}\{L(\mu(x), Y)\} \geq \mathbb{E}_{\mathrm{do}(X = x)}\{L(\mu^*(X), Y)\} = \mathbb{E}\{L(\mu^*(X), Y)\}.
\]
Here $x \in \mathbb{R}$ is arbitrary, and the second inequality holds $L\{\mu^*(x), Y\} = L(0, \varepsilon_Y)$ and because $0 = T\{\mathcal{L}(\varepsilon_Y)\}$; equality holds if and only if $\mu(x) = \mu^*(x)$ due to strict consistency of $L$.
\end{proof}

\subsection{Proof of Theorem 1}
\begin{proof}
To obtain a contradiction, suppose that there exists a strictly proper scoring rule $S$ such that $\pi^*_{y|x}$ is $(S,\cP)$-invariant, for which we assume without any loss of generality that the risk is constant equal to $c = 0$ by subtracting $c$ form $S$. We have
\[
    0 = \mathbb{E}_{\mathrm{do}\{X = (x_1, x_2)\}}(S[\mathcal{N}\{\mu^*(X_1), \sigma^*(X_2)^2\}, Y]) = \mathbb{E}_{Y \sim \mathcal{N}\{\mu^*(x_1), \sigma^*(x_2)^2\}}(S[\mathcal{N}\{\mu^*(x_1), \sigma^*(x_2)^2\}, Y]).
\]
Since $\mu^*$ and $\sigma^*$ are surjective, this yields
\[
    0 = \mathbb{E}_{Y \sim \mathcal{N}(\mu, \sigma^2)}[S\{\mathcal{N}\{\mu, \sigma^2), Y\}]
\]
for all fixed $\mu \in \mathbb{R}$, $\sigma > 0$. Because $S$ is strictly proper, we also know
\begin{equation} \label{eq:score_positive}
    \mathbb{E}_{Y \sim \mathcal{N}\{m, s^2\}}[S(\mathcal{N}m, s^2), Y)] = 0 < \mathbb{E}_{Y \sim \mathcal{N}(m, s^2)}[S\{\mathcal{N}(\mu, \sigma^2), Y\}]
\end{equation}
for all $m \in \mathbb{R}$ and $s > 0$ such that $m \neq \mu$ or $s \neq \sigma$. Let $V = Z + Z'$, where $Z, Z' \sim \mathcal{N}(0,1)$ independent, so $V \sim \mathcal{N}(0,2)$. Then we have
\[
    \mathbb{E}[S(\mathcal{N}(0,2), V)] = \mathbb{E}(\mathbb{E}[S\{\mathcal{N}(0,2), Z + Z'\} \mid Z]) > 0,
\]
because $\mathbb{E}[S\{\mathcal{N}(0,2), Z + Z' \mid Z] > 0$ a.s.~by \eqref{eq:score_positive}, since $Z + Z' \sim \mathcal{N}(Z, 1)$ conditional on $Z$. This contradicts $\mathbb{E}[S\{\mathcal{N}(0,2), V\}] = 0$.
\end{proof}

\subsection{Proofs for Example 1}
The following lemma is applied several times in the proof.
\begin{lemma} \label{lem:gaussian}
If $X \sim \mathcal{N}_n(0, \Sigma)$ and $\theta \in \mathbb{R}^n$, then 
\[
    \mathbb{E}\{h(X)\exp(\theta^{\top}X)\} = \mathbb{E}\{h(Y)\}\exp(\theta^{\top}\Sigma\theta/2),
\]
where $h\colon \mathbb{R}^n \rightarrow \mathbb{R}$ is any function such that the expected value exists, and $Y \sim \mathcal{N}(\Sigma\theta, \Sigma)$.
\end{lemma}
\begin{proof}[Proof of Lemma \ref{lem:gaussian}]
Combining the factor $\exp(\theta^{\top}X)$ with the multivariate Gaussian density gives
\[
    h(x)c(\Sigma)\exp(\theta^{\top}x)\exp(-x^{\top}\Sigma^{-1}x/2) = h(x)c(\Sigma)\exp(-(x-\Sigma\theta)^{\top}\Sigma^{-1}(x-\Sigma\theta)/2)\exp\{\theta^{\top}\Sigma\theta/2\},
\]
where $c(\Sigma)$ is the normalization constant.
\end{proof}

For the proof, we simplify the notation. All random variables and quantities in this proof are not related to those in the main text, e.g.~a random variable $Y$ in this proof is not related to the $Y$ in the structural causal model.

Let $(U, V)$ be a Gaussian random vector with zero mean and covariance matrix $\Sigma \in \mathbb{R}^{(d+1)\times(d+1)}$ such that $\Sigma_{11} = 1$. Let furthermore $\Gamma \in \mathbb{R}^{d\times d}$ be an invertible matrix, and $\theta, \eta \in \mathbb{R}^d$. We need to compute expectations of random variables of the following types,
\begin{align*}
    A & = (\theta^{\top}\Gamma V)^2\exp(\eta^{\top}\Gamma V), \\
    B & = U^2\exp(\theta^{\top}\Gamma V), \\
    C & = U \cdot \theta^{\top}\Gamma V\exp(\eta^{\top}\Gamma V).
\end{align*}

In the following, let $M = (\Sigma_{ij})_{i,j=2}^{d+1}$, which is the covariance matrix of $V$. For $A$, we can use that $W = \Gamma V$ is Gaussian with mean $0$ and covariance matrix $\Gamma M \Gamma^{\top}$. By Lemma \ref{lem:gaussian}, we have
\[
    \mathbb{E}\{(\theta^{\top}W)^2\exp(\eta^{\top}W)\} = \mathbb{E}\{(\theta^{\top}Y)^2\}\exp(\eta^{\top}\Gamma M \Gamma^{\top}\eta/2),
\]
where $Y \sim \mathcal{N}(\Gamma M \Gamma^{\top}\eta, \Gamma M \Gamma^{\top})$. The expectation of $(\theta^{\top}Y)^2$ is $\theta^{\top}\Gamma M \Gamma^{\top}\theta + (\theta^{\top}\Gamma M \Gamma^{\top}\eta)^2$, so
\[
    \mathbb{E}(A) = \exp(\eta^{\top}\Gamma M \Gamma^{\top}\eta/2)\{\theta^{\top}\Gamma M \Gamma^{\top}\theta + (\theta^{\top}\Gamma M \Gamma^{\top}\eta)^2\}.
\]
For the variable $B$, \revision{let $\xi = (0, \Gamma^\top \theta)^\top$, $e_{1}$ denote the first standard basis vector, $s = (\Sigma_{i1})_{i=2}^{d+1}$, $X = (U, V^\top)^\top$, and $Y \sim \mathcal{N}(\Sigma \xi, \Sigma)$.  Then, 
\begin{align*}
    \mathbb{E}(B)
    = \mathbb{E}\{(e_{1}^\top X)^{2} \exp(\xi^\top X)\}
    = \mathbb{E}\{(e_{1}^\top Y)^{2}\} \exp(\xi^\top \Sigma \xi / 2)
    = \{(s^\top \Gamma^\top \theta)^{2} + 1\} \exp(\theta^\top \Gamma M \Gamma^\top \theta / 2).
\end{align*}
The last equality follows from the fact observation that $e_{1}^\top Y \sim \mathcal{N}(s^\top \Gamma^\top \theta, 1)$.}

Finally, for $C$, 
define $Z = \Gamma V$. Then the vector $(U, Z^{\top})$ has mean $0$ and covariance matrix 
\[
	D = \left(\begin{array}{cc}
		1 & 0 \\
		0 & \Gamma
	\end{array}\right)
	\Sigma \left(\begin{array}{cc}
		1 & 0 \\
		0 & \Gamma^{\top}
	\end{array}\right)
	= \left(\begin{array}{cc}
		1 & (\Gamma s)^{\top} \\
		\Gamma s & \Gamma M\Gamma^{\top}
	\end{array}\right).
\]
Lemma \ref{lem:gaussian} implies that
\[
	\mathbb{E}(C) = \mathbb{E}(U \theta^{\top}K)\exp(\eta^{\top}\Gamma M \Gamma^{\top}\eta/2),
\]
where $(U,K^{\top}) \sim \mathcal{N}_{d+1}\{(0,\eta^{\top})^{\top}, D\}$. Since $\mathbb{E}(U) = 0$, $\mathbb{E}(U \theta^{\top}K)$ equals the covariance of $U$ and $\theta^{\top}K$. The covariance matrix of $(U,K^{\top})$ is
\[
\left(\begin{array}{cc}
	1 & 0 \\
	0 & \theta^{\top}
\end{array}\right)
D \left(\begin{array}{cc}
	1 & 0 \\
	0 & \theta
\end{array}\right)
= \left(\begin{array}{cc}
	1 & s^{\top}\Gamma^{\top}\theta \\
	\theta^{\top}\Gamma s & \eta\Gamma M \Gamma^{\top}\theta^{\top}
\end{array}\right),
\]
so we obtain
\[
	\mathbb{E}[C] = \theta^{\top}\Gamma s\exp(\eta^{\top}\Gamma M \Gamma^{\top}\eta/2)
\]

In the example, we have the following parameters
\begin{align*}
	& A \colon \theta = \beta - b, \ \eta = -2g, \\
	& B \colon \theta = 2(\gamma - g), \\
	& C \colon \theta = 2(\beta - b), \ \eta = \gamma - 2g.
\end{align*}
Identifiability except for the third case directly follows from the formula for the LogS. The proof for the third case is by contradiction. Assume that there exist parameters $b', g'$ with $b' \neq \beta$ or $g' \neq \gamma$ for which the risk is constant and minimal. If $b' = \beta$, then the minimum is achieved for $g = \gamma$, which contradicts $g' \neq \gamma$. Hence $b' \neq \beta$. If $(\beta - b')^{\top}\Gamma^e \Sigma_{YX} = 0$ for $e = e_1, \dots, e_d$, then by linear independence we have $\beta = b'$, which again yields a contradiction. Hence $(\beta -b')^{\top}\Gamma^e \Sigma_{YX} \neq 0$ for some environment $e$. If $(\beta -b')^{\top}\Gamma^e \Sigma_{YX} > 0$ for some $e$, then the risk is again greater than that of $b = \beta$, $g = \gamma$. If $(\beta -b')^{\top}\Gamma^e \Sigma_{YX} < 0$ for $e = e_1, \dots, e_d$, then $(\beta -b')^{\top}\Gamma^{e_{d+1}} \Sigma_{YX} > 0$ by definition of $e_{d+1}$. Hence the risk is again greater than that of $b = \beta$, $g = \gamma$.

\revision{\subsection{Proof of Proposition 3}
\begin{proof}
By assumption, the distribution of $\varepsilon_Y$
\[
    \varepsilon_Y = \frac{Y^{\mathrm{int}}-\mu^*(X^{\mathrm{int}})}{\sigma^{*}(X^{\mathrm{int}})}
\]
does not change under interventions. We also have
\[
    \ell^*_{X^{\mathrm{int}}}(\alpha)= \mu^*(X^{\mathrm{int}}) + \sigma^*(X^{\mathrm{int}}) q_{\mathrm{\alpha}/2},
\]
where $q_{\mathrm{\alpha}/2}$ is the $\alpha/2$-quantile of the distribution of $\varepsilon_Y$; the same holds for $u^*_{X^{\mathrm{int}}}(\alpha)$, with the $(1-\alpha/2)$-quantile $q_{1-\alpha/2}$. Hence,
\begin{align*}
    \mathbb{P}\{Y^{\mathrm{int}} \in [\ell^*_{X^{\mathrm{int}}}(\alpha), u^*_{X^{\mathrm{int}}}(\alpha)]\} = \mathrm{P}\{\varepsilon_Y \in [q_{\alpha/2}, q_{1-\alpha/2}]\}
    \geq 1-\alpha.
\end{align*}
\end{proof}
}

\subsection{Proof of Theorem 2}

\begin{lemma} \label{lem:population_convergence}
Consider the setting of Theorem 2.  Let 
\begin{align*}
    \theta(\lambda) = \argmin_{\theta \in \Theta} \mathcal{R}_{\lambda}(\theta),
\end{align*}
where 
\begin{align*}
    \mathcal{R}_{\lambda}(\theta) = \sum_{e=1}^E\omega^e\mathcal{R}^e(\theta) + \lambda D\{\mathcal{R}^1(\theta), \dots, \mathcal{R}^E(\theta)\}.
\end{align*}
Then, 
\[
    \lim_{\lambda\rightarrow\infty}\theta(\lambda) = \theta^*.
\]
\end{lemma}
\begin{proof}[Lemma \ref{lem:population_convergence}]

For an arbitrary sequence $\lambda_{n} \to \infty$, let $\theta_{n} = \theta(\lambda_{n})$.  Since $\Theta$ is compact, every subsequence $(n_{k})_{k=1}^{\infty} \subseteq \mathbb{N}$ admits a further $(n_{k_{\ell}})_{\ell=1}^{\infty} \subseteq (n_{k})_{k=1}^{\infty}$ such that $\theta_{n_{k_{\ell}}} \to \theta_0$ for some $\theta_0 \in \Theta$.  Now, if $\theta^{\ast} \neq \theta_{0}$, then either $\theta_{0}$ is not invariant or $\mathcal{R}^{1}(\theta_0) > \mathcal{R}^{1}(\theta^{\ast})$.  In the first case, if $\theta_{0}$ is not invariant, then there exists $\epsilon > 0$ such that $D\{\mathcal{R}^{1}(\theta_0), \dots, \mathcal{R}^{E}(\theta_0)\} \geq \epsilon$.  As the mapping $\theta \mapsto D\{\mathcal{R}^{1}(\theta_0), \dots, \mathcal{R}^{E}(\theta_0)\}$ is continuous and $\theta_{n_{k_{\ell}}} \to \theta_{0}$, there exists an $N_\epsilon \in \mathbb{N}$ such that $D\{\mathcal{R}^{1}(\theta_{n_{k_{\ell}}}), \dots, \mathcal{R}^{E}(\theta_{n_{k_{\ell}}})\} \geq \epsilon / 2$ for all $n_{k_{\ell}} \geq N_\epsilon$, implying that 
\begin{align*}
    \liminf_{\ell \to \infty} \mathcal{R}_{\lambda_{n_{k_{\ell}}}}(\theta_{n_{k_{\ell}}}) \geq \min_{\theta \in \Theta} \Big\{\frac{1}{E}\sum_{e=1}^E\omega^e\mathcal{R}^e(\theta)\Big\} + \liminf_{\ell \to \infty}\lambda_{n_{k_{\ell}}} \frac{\epsilon}{2} \to \infty.
\end{align*}
Hence, $\theta_{n_{k_{\ell}}}$ is not a minimizer of $\mathcal{R}_{\lambda}(\theta)$ as $\mathcal{R}_{\lambda_{n_{k_{\ell}}}}(\theta^{\ast})$ is always finite.  Thus, we conclude that $\theta_{0}$ is invariant.

Now, for the second case, let $\delta = \mathcal{R}^{1}(\theta_0) - \mathcal{R}^{1}(\theta^{\ast}) > 0$.  We observe that $\mathcal{R}^{1}$ is continuous as $S$ is a continuous proper scoring rule.  Since $\mathcal{R}^{1}$ is continuous, there exists an $N_\delta \in \mathbb{N}$ such that $\mathcal{R}^{1}(\theta_{n_{k_{\ell}}}) \geq \mathcal{R}^{1}(\theta_0) - \delta / 2$ for all $\ell \geq N_{\delta}$.  Then, as $\theta^{\ast}$ is invariant, we note that 
\begin{align*}
    \mathcal{R}_{\lambda_{n_{k_{\ell}}}}(\theta_{n_{k_{\ell}}}) - \mathcal{R}_{\lambda_{n_{k_{\ell}}}}(\theta^{\ast})
    \geq \mathcal{R}^{1}(\theta_{n_{k_{\ell}}}) - \mathcal{R}^{1}(\theta^{\ast})
    \geq \delta / 2
\end{align*}
for all $\ell \geq N_{\delta}$.  This contradicts the assumption that $\theta_{n_{k_{\ell}}}$ is a minimizer, which concludes the proof.

\end{proof}

\begin{proof}[Theorem 2]

We start by showing that 
\begin{align*}
    \sup_{\theta \in \Theta}|\mathcal{R}_{\lambda_n}(\theta) - \hat{\mathcal{R}}_{\lambda_n}(\theta)| \overset{P}{\rightarrow} 0.
\end{align*}
Indeed, since $\mathcal{R}^{1}, \dots, \mathcal{R}^{E}$ are continuous, Theorem 1.1 of \cite{Ledoux1988} with the Banach space of continuous functions on the compact set $\Theta$ and the supremum norm implies 
\begin{align*}
    \sup_{\theta\in\Theta}|\hat{\mathcal{R}}^k(\theta) - \mathcal{R}^k(\theta)|[n\log\{\log(n)\}]^{1/2} = O_P(1).
\end{align*}
Combining the above with Assumption 6 yields 
\begin{align*}
    \sup_{\theta \in \Theta}|\mathcal{R}_{\lambda_n}(\theta) - \hat{\mathcal{R}}_{\lambda_n}(\theta)| \overset{P}{\rightarrow} 0.
\end{align*}
Next, letting $\theta(\lambda)$ be defined as in Lemma \ref{lem:population_convergence}, we have 
\begin{align*}
    \hat{\mathcal{R}}_{\lambda}\{\hat{\theta}(\lambda)\} 
    \leq \hat{\mathcal{R}}_{\lambda}\{\theta(\lambda)\} 
    \leq \Big| \hat{\mathcal{R}}_{\lambda}\{\theta(\lambda)\} - \mathcal{R}_{\lambda}\{\theta(\lambda)\} \Big| + \mathcal{R}_{\lambda}\{\theta(\lambda)\}
    = \mathcal{R}_{\lambda}\{\theta(\lambda)\} + o_{P}(1)
\end{align*}
and 
\begin{align*}
    \mathcal{R}_{\lambda}\{\theta(\lambda)\} 
    \leq \mathcal{R}_{\lambda}\{\hat\theta(\lambda)\} 
    \leq \Big| {\mathcal{R}}_{\lambda}\{\theta(\lambda)\} - \hat{\mathcal{R}}_{\lambda}\{\hat\theta(\lambda)\} \Big| + \hat{\mathcal{R}}_{\lambda}\{\hat\theta(\lambda)\}
    = \hat{\mathcal{R}}_{\lambda}\{\hat\theta(\lambda)\} + o_{P}(1).
\end{align*}
Together with Lemma \ref{lem:population_convergence}, this implies that 
\begin{align*}
    \Big| \mathcal{R}_{\lambda}\{\hat\theta(\lambda)\}  - \mathcal{R}_{\lambda}(\theta^\ast) \Big| = o_{P}(1).
\end{align*}
Hence, we have $D\{\mathcal{R}^{1}(\hat{\theta}_{n}\}, \dots, \mathcal{R}^{E}(\hat{\theta}_{n})) \overset{P}{\rightarrow} 0$ since $\lambda_{n} \to \infty$.  Thus, every subsequence $\hat\theta_{n_{k}}$ of $\hat{\theta}_{n}$ has a further subsequence $\hat\theta_{n_{k_{\ell}}}$ that converges to an invariant parameter $\theta_{0} \in \Theta$.  Since $\mathcal{R}^{1}$ is continuous, it follows that $\theta_0 = \theta^\ast$ from the proof of Lemma \ref{lem:population_convergence}, which finishes the current proof.
\end{proof}

\subsection{Proof of Lemma 3}
\begin{proof}
Let $0 \leq \lambda_1 < \lambda_2$ and define
\[
    A_j = \frac{1}{E}\sum_{k=1}^E\mathcal{R}^k(\theta_{\lambda_j}), \ B_j = D\{\mathcal{R}^1\{\theta_{\lambda_j}\}, \dots, \mathcal{R}^E(\theta_{\lambda_j})\}, \ j = 1, 2.
\]
Then we have
\[
    A_1 + \lambda_2 B_1 \geq A_2 + \lambda_2 B_2, \ A_2 + \lambda_1 B_2 \geq A_1 + \lambda_1 B_1,
\]
which implies
\[
    \lambda_2(B_1 - B_2) \geq A_2 - A_1 \geq \lambda_1(B_1 - B_2).
\]
Since $\lambda_2 > \lambda_1$, this can only be true if $B_1 \geq B_2$.
\end{proof}

\section{Experimental details} \label{sec:experimental_details}
For the \texttt{R} implementation of our methdos with the Gaussian heteroscedastic linear model, we apply the package \texttt{GA} \citep{Scrucca2013} for optimization. The algorithms in this package do not assume convexity, and require the specification of a compact box for parameter search, which we to $[-5, 5]$ for all parameters in the simulation and case study. 

Our Python implementations of IPP with NN models are based on the machine learning framework \texttt{PyTorch}~\citep{paszke2019pytorch}. We take the same model architecture as in \citet[Figure 17]{shen2023engression} with 3 layers and 400 neurons per layer; at each layer, the input vector is concatenated with an independently sampled 400-dimensional standard Gaussian noise. We adopt the Adam optimizer~\citep{kingma2014adam} with a learning rate of $1\times10^{-2}$ and train all the models for 1k iterations. We keep all the hyperparameters the same for V-REx. For each $\lambda$, we repeat the experiments with random initializations for 10 times and take the average of their evaluation metrics.

\revision{Our implementation for conformal prediction is based on the code by \citet{Ai2024}, available on \url{https://github.com/zhimeir/finegrained-conformal-paper}, taking linear regression to obtain nonconformity scores and logistic regression as method for obtaining the weights in the weighted variants of conformal prediction. For the train-test split we randomly split the training environments into two halves. For the debiased weighted conformal prediction, we follow Algorithm 2 of \citet{Ai2024} and take, separately for each environment, half of the test data covariates to estimate the weight function, and predict on the other half. We slightly adapted the code by \citet{Ai2024} so that it takes the maximum of the nonconformity scores instead of infinity in their definition (2) to obtain intervals that can be compared to other methods in such cases, even if the theoretical properties might be unclear. The range of the penalty parameter $\rho$ for conformal prediction is chosen between $0$ and $0.04$, which is the same as in \citet{Ai2024}.}

\section{Additional figures} \label{sec:add_figures}
\begin{figure}
    \centering
    \includegraphics[width=0.8\textwidth]{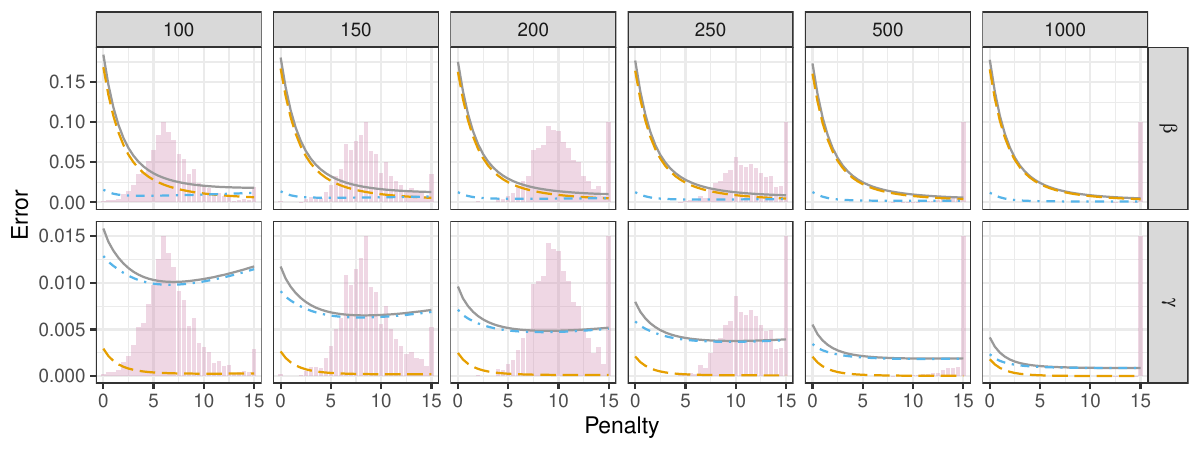}
    \caption{As Fig.~\ref{fig:parameter_error} but for estimation with SCRPS instead of LogS. \label{fig:parameter_error_scrps}}
\end{figure}

\begin{figure}
    \centering
    \includegraphics[width=0.8\textwidth]{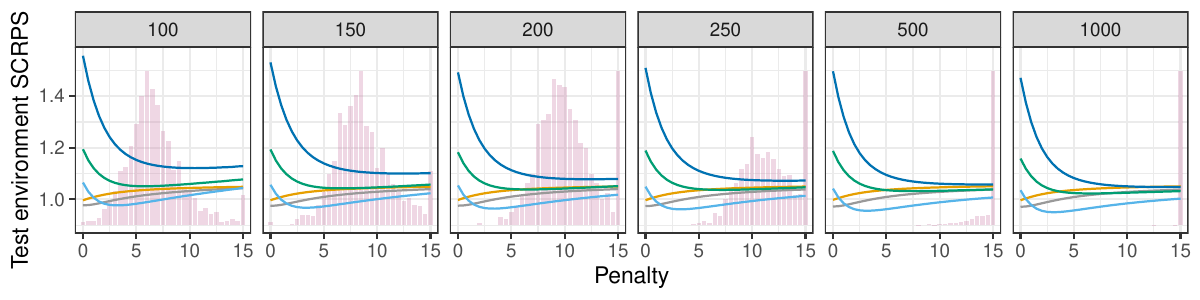}
    \caption{As Fig.~\ref{fig:interventions} but for estimation and evaluation with SCRPS instead of LogS. \label{fig:interventions_scrps}}
\end{figure}

\begin{figure}
    \centering
    \includegraphics[width=0.8\textwidth]{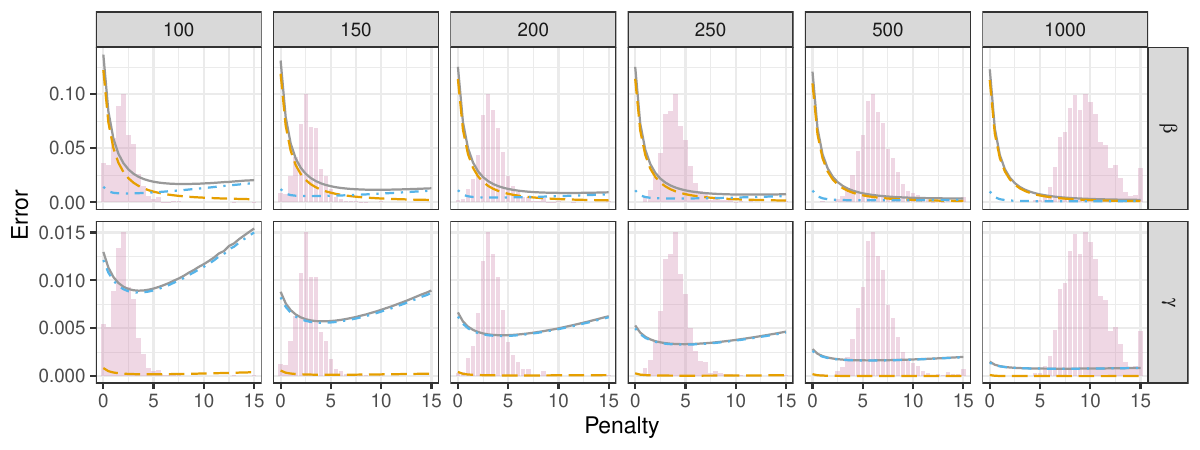}
    \caption{As Fig.~\ref{fig:parameter_error} but with $\alpha = 0.01$ for choosing the penalty parameter. \label{fig:parameter_error_alpha0.01}}
\end{figure}

\begin{figure}
    \centering
    \includegraphics[width=0.8\textwidth]{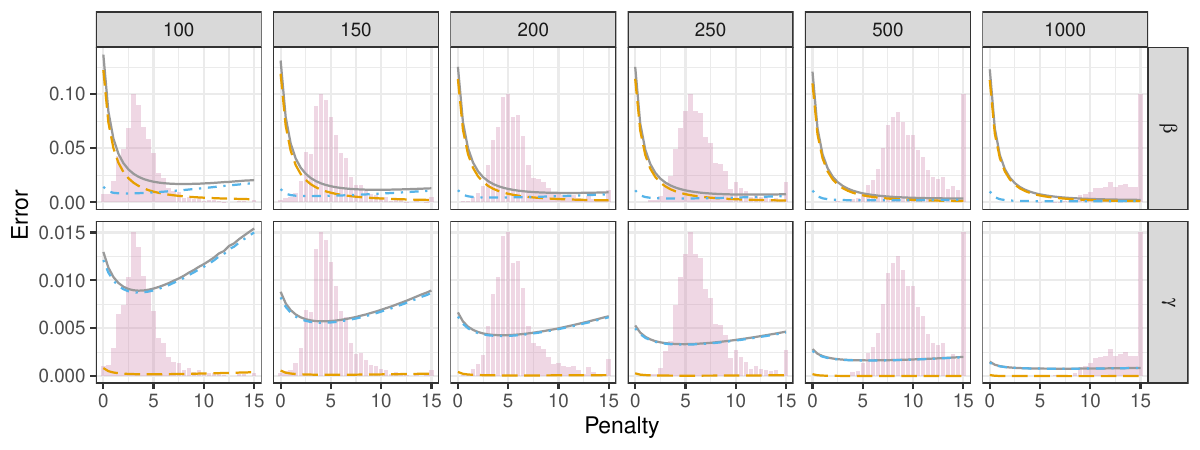}
    \caption{As Fig.~\ref{fig:parameter_error} but with $\alpha = 0.1$ for choosing the penalty parameter. \label{fig:parameter_error_alpha0.1}}
\end{figure}

Figures \ref{fig:parameter_error_scrps} and \ref{fig:interventions_scrps} are the same as Figs.~\ref{fig:parameter_error} and \ref{fig:interventions} but with SCRPS instead of LogS, and Figs.~\ref{fig:parameter_error_alpha0.01} and \ref{fig:parameter_error_alpha0.1} are as Fig.~\ref{fig:parameter_error} but $\alpha = 0.01$ and $\alpha = 0.1$ for the selection of $\lambda$.

\begin{figure}
    \centering
    \includegraphics[width=0.8\textwidth]{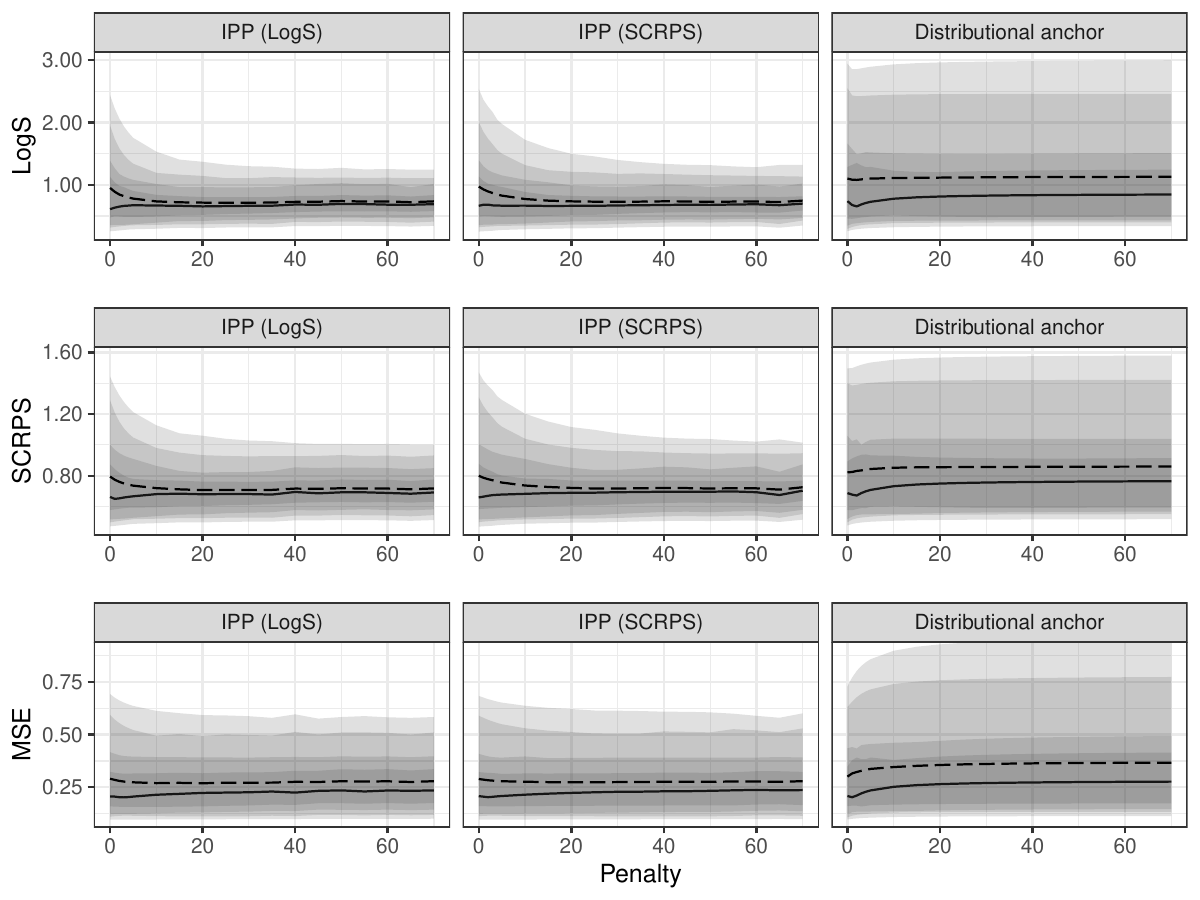}
    \caption{Comparison of IPP without the observational training environment to distributional anchor regression, on LogS, SCRPS, and MSE. \label{fig:single_cell_without_observational}}
\end{figure}

\begin{figure}
    \centering
    \includegraphics[width=0.9\textwidth]{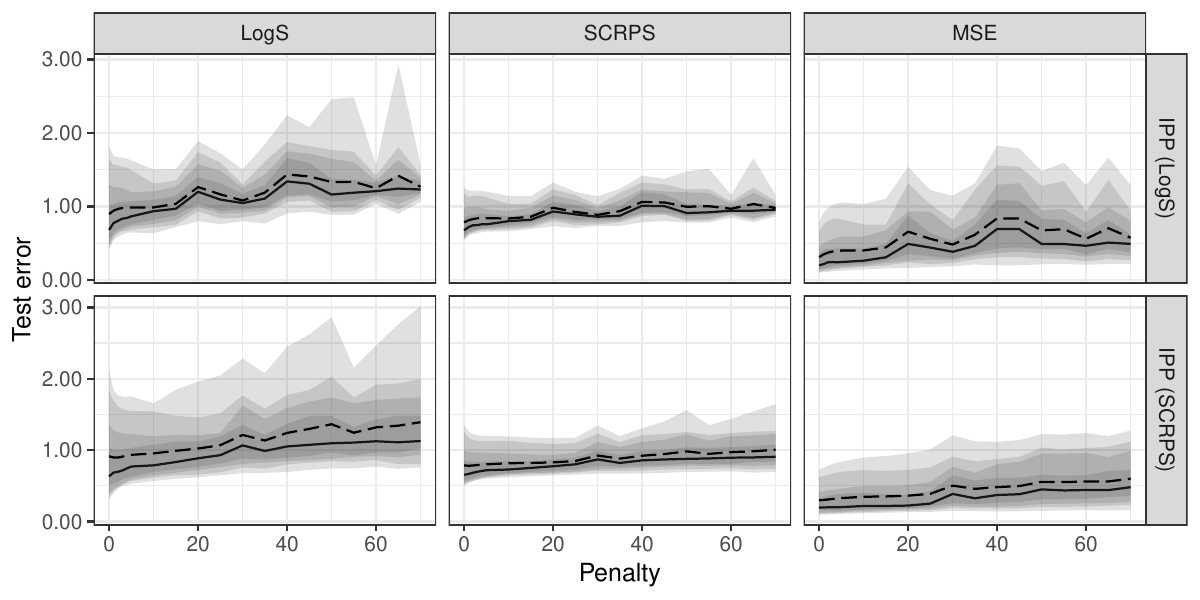}
    \caption{Scores of IPP when the environment with interventions on the response gene is included in the training data. \label{fig:sincle_cell_rvar}}
\end{figure}

\begin{figure}
    \centering
    \includegraphics[width=\textwidth]{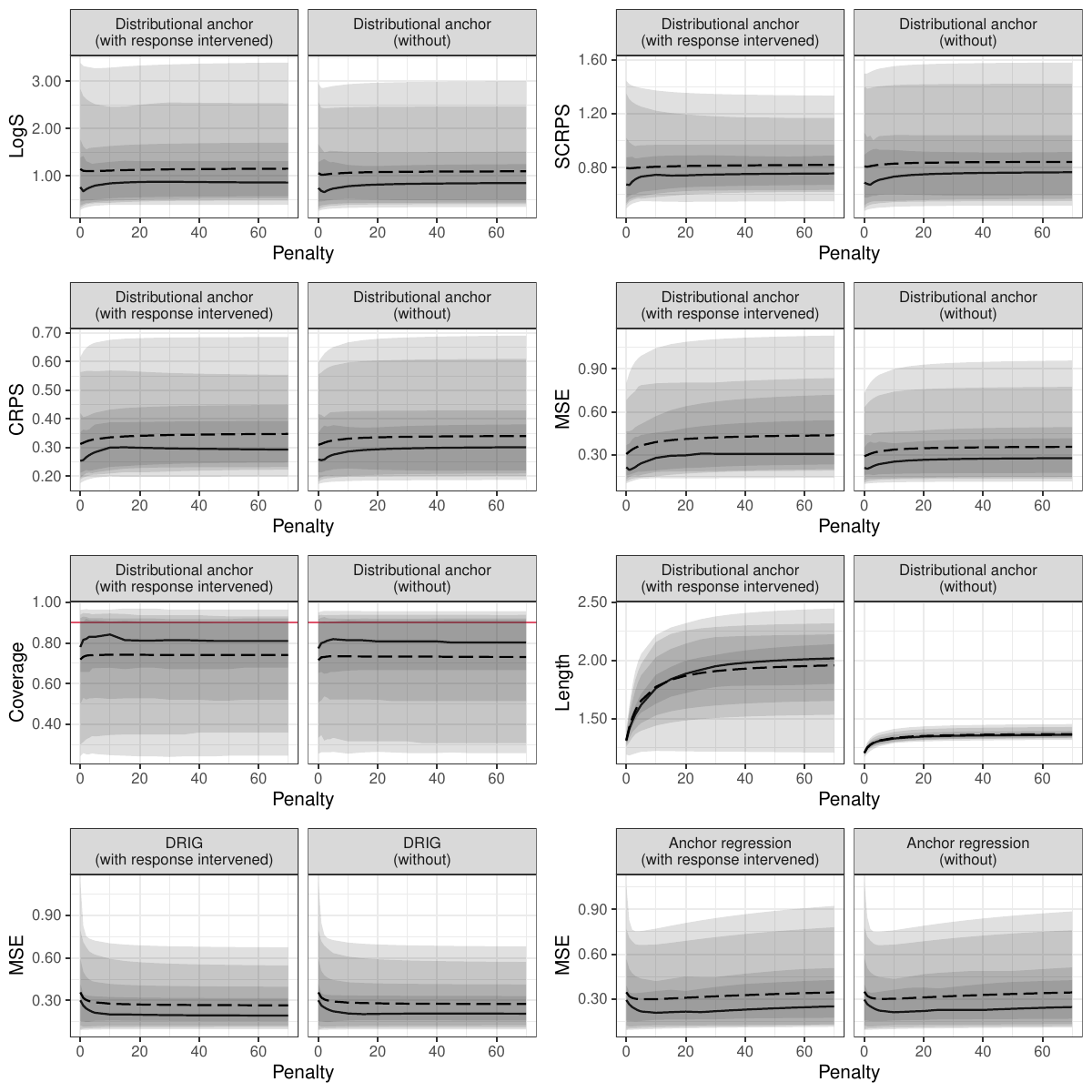}
    \caption{Comparison of distributional anchor regression, DRIG, and anchor regression with and without the environment with interventions on the response variable in the training data. \label{fig:anchor_with_response_interv}}
\end{figure}

\begin{figure}
    \centering
    \includegraphics[width=0.8\textwidth]{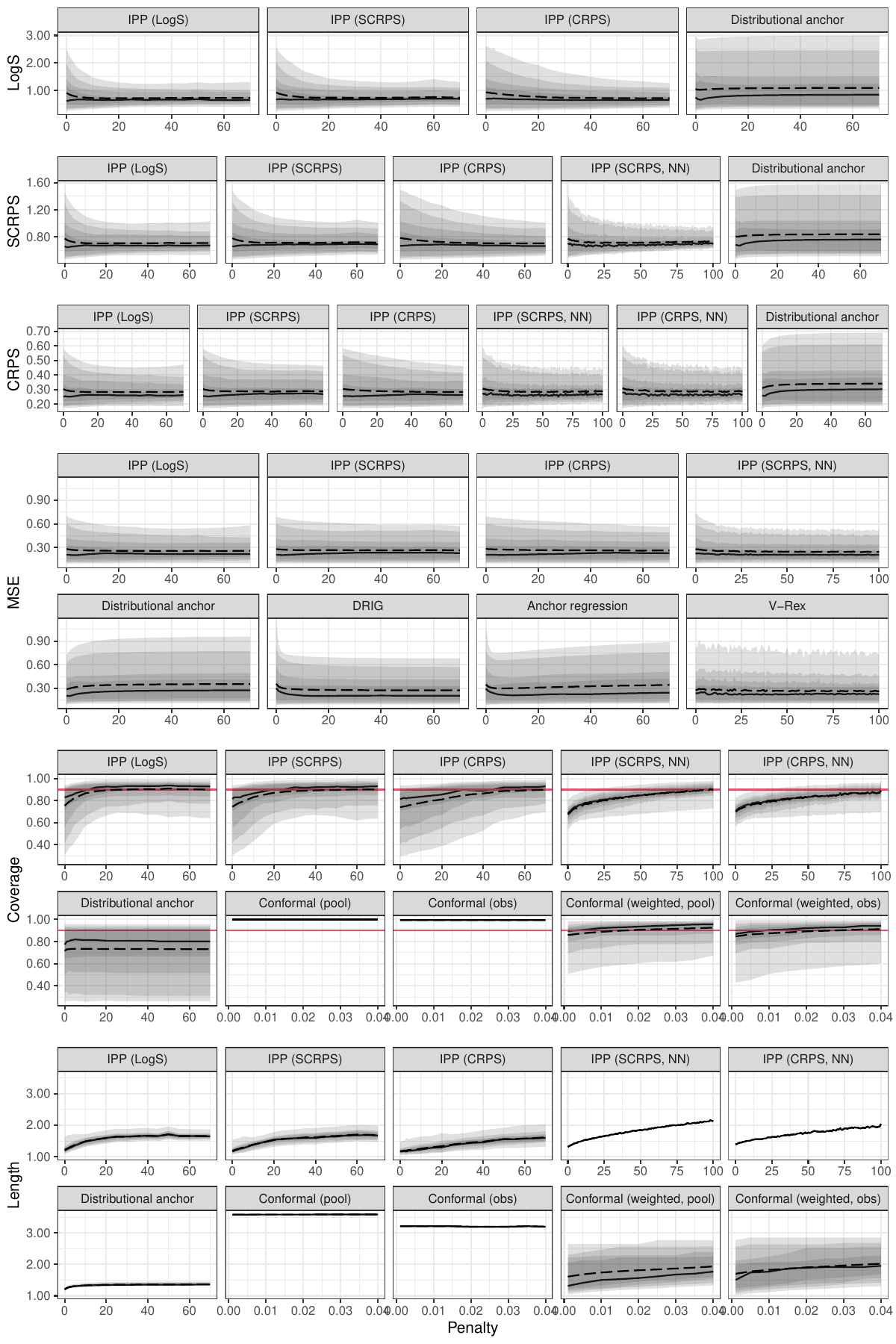}
    \caption{As Fig.~\ref{fig:single_cell_test} but including additionally the CRPS and further variants of the prediction methods. \label{fig:expanded_comparison}}
\end{figure}

In Fig.~\ref{fig:single_cell_without_observational} we compare the implementations of IPP with the heteroscedastic linear model against distributional anchor regression when the former exclude the observational environment from the training data. As in Section \ref{sec:single_cell}, the variants of IPP achieve a smaller mean error and less variability between the test environments. In Fig.~\ref{fig:sincle_cell_rvar} we depict the scores of IPP when the training data includes the environment where the response gene is intervened on. The theoretical setup for IPP does not allow interventions directly on the response variable, and one usually cannot expect an invariant risk under such interventions; it is therefore not surprising that stronger penalization has a negative effect on the risks in this case. \revision{Figure \ref{fig:anchor_with_response_interv} compares distributional anchor regression, anchor regression, and DRIG with and without including the environment where the response gene is intervened on in the training data. Apart from the length of the distributional anchor regression prediction intervals, which increases when one includes the additional environment, there are no noticeable differences. Figure \ref{fig:expanded_comparison} is an extended version of Fig.~\ref{fig:single_cell_test}, where the probabilistic predictions are also evaluated with respect to CRPS, and which includes the following additional methods: the parametric and neural network variants of IPP trained by CRPS minimization, and conformal prediction trained on the pooled training environments instead of only the observational environments. For conformal prediction, taking the pooled training environments does no improve the worst-case coverage and only has a slight effect on the prediction interval lengths. The IPP variants trained on CRPS generally have good performance, and penalizing towards invariant CRPS also leads to less variation in the test SCRPS and MSE and improves the prediction interval coverage.}

\clearpage
\section{Comparison on semi-real data} \label{sec:sim2}

\subsection{Data generation}
\begin{figure}
    \centering
    \includegraphics[width=0.6\textwidth]{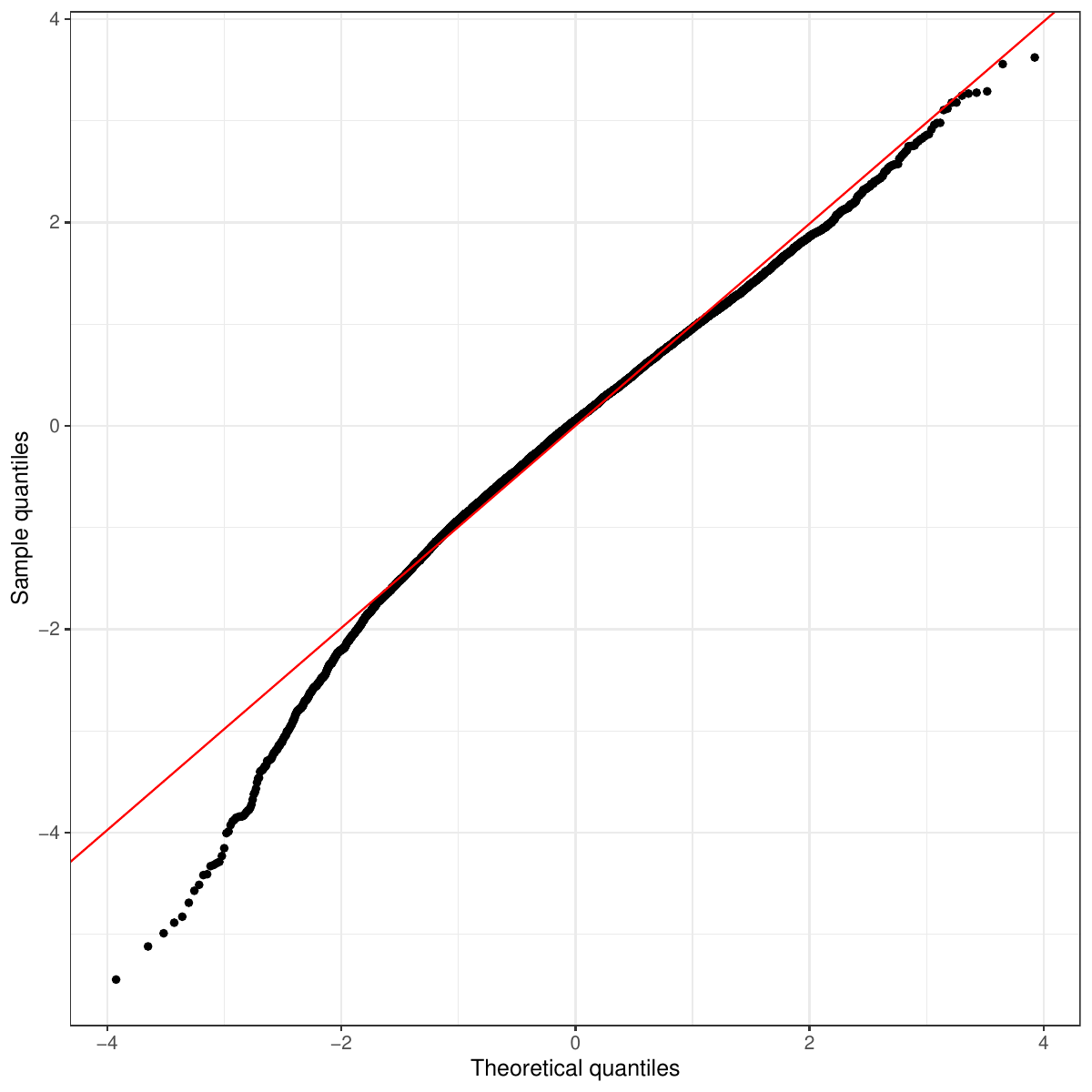}
    \caption{Normal QQ-plot of the distribution of $\varepsilon_Y$ in the data analysis from Section \ref{sec:sim2}. \label{fig:simulation2qq}}
\end{figure}

In this section we analyze IPP in a simulation-based modification of the single cell dataset from Section \ref{sec:single_cell}. We generate observations
\begin{equation} \label{eq:sim2_data_generation}
    Y^e \leftarrow \beta_0 + \beta^{\top}X^{e} + \beta_H^{\top}H^e + \exp\left(\gamma_0 + \gamma^{\top} X^e + \gamma_H^{\top} H^e\right) \varepsilon_Y,
\end{equation}
where the variables are constructed as follows. We separate the $10$ genes into $3$ groups. The gene with identifier {\tt ENSG00000173812} is the response variable $Y$, which is the same gene as in Section \ref{sec:single_cell}. Those with identifiers {\tt ENSG00000187514}, {\tt ENSG00000075624}, {\tt ENSG00000147604}, and {\tt ENSG00000110700} are later on treated as hidden variables $H_1, \dots, H_4$; and the remaining $5$ genes are observed covariates $X_1, \dots, X_5$. Let $y_i$, $h_i = (h_{1,i}, \dots, h_{4, i})$, and $x_i = (x_{1,i}, \dots, x_{5,i})$, $i = 1, \dots, N$, denote the gene expressions in the observational data. In a first step, we perform a linear regression of the $y_i$ on $h_i$ and $x_i$, which yields $\beta_0$, $\beta$, and $\beta_H$. Let $r_i = y_i - \beta_0 - \beta^{\top}x_i - \beta_H^{\top}h_i$ be the residuals of the regression. To obtain the scale parameters, we regress $\log(|r_i|)$ the on $x_i$ and $h_i$, and define
\begin{align*} 
    \tilde{\varepsilon}_i = \frac{r_i}{\exp(\gamma_0 + \gamma^{\top}x_i + \gamma_H^{\top}h_i)}, \ i = 1, \dots, N.
\end{align*}
The variable $\varepsilon_Y$ is then drawn of the empirical distribution of $\tilde{\varepsilon}_1, \dots, \tilde{\varepsilon}_N$ standardized to zero mean and standard deviation $1$. The resulting distribution is asymmetric --- it has a longer lower tail --- and has heavier tails than Gaussian; see the normal QQ-plot in Fig.~\ref{fig:simulation2qq}. Table \ref{tab:single_cell_info} gives the parameters values for \eqref{eq:sim2_data_generation} and the number of observations in each environment.

\begin{table}[th]
    \centering
    \small
    \begin{tabular}{ccccc}
        Object & Treatment & Location & Scale & Samples\\[0.1em]
        ENSG00000125691 & observed & -0.003 & -0.073 & 101\\
        ENSG00000108518 & observed & 0.048 & 0.019 & 122\\
        ENSG00000067225 & observed & -0.002 & 0.027 & 159\\
        ENSG00000133112 & observed & 0.008 & 0.029 & 552\\
        ENSG00000172757 & observed & 0.069 & -0.069 & 139\\
        ENSG00000110700 & hidden & 0.041 & -0.080 & 121\\
        ENSG00000147604 & hidden & 0.037 & 0.014 & 116\\
        ENSG00000075624 & hidden & 0.036 & -0.118 & 229\\
        ENSG00000187514 & hidden & 0.049 & -0.090 & 180\\
        ENSG00000173812 & response & - & - & 146 \\[0.25em]
        Intercept & - & 1.842 & -0.706 & -\\
        Observational & - & - & - & 11485\\[0.5em]
    \end{tabular}
    \caption{Summary of genes and data availability in the single cell data. The column `Object' refers to gene identifiers, or the intercept parameters $\beta_0$, $\gamma_0$, in \eqref{eq:sim2_data_generation}, or the environment, for the observational data. Location and scale parameters refer to $\beta, \beta_H$ and $\gamma, \gamma_H$, respectively. The last columns gives the number of available samples in intervened environments.}
    \label{tab:single_cell_info}
\end{table}

As environments $e$, we take the observational environment as $e = 1$, and the remaining $9$ environments with interventions on each single of the genes $X_i$, $i = 1, \dots, 5$, and $H_i$, $i = 1, \dots, 4$. To generate training environments, we sample $1000$ observations $(X^e, H^e)$ from the observational environments and $80$ from each of the environments with interventions on $X_1, \dots, X_5$ with replacement, which yields a training dataset consisting of $6$ environments and total $1400$ observations. The response variable is generated according to \eqref{eq:sim2_data_generation}, with $\varepsilon$ drawn independently of $(X^e, H^e)$. Test environments are generated in the same way, with $80$ observations for each environment, inculding those with interventions on $H_1, \dots, H_4$, which gives $10$ test environments. This data generation process, for both training and test data, is then repeated $500$ times, and we compute the average risk of different methods over the test environments.

When comparing IPP to competitors, we treat $H_1, \dots, H_4$ as unobserved variables and only use $X_1, \dots, X_5$ in regressions. This creates data similar to the real data, with the same interventions on the covariates, but the observations generated according to the known mechanism \eqref{eq:sim2_data_generation}. In this setting, the parametric IPP is misspecified: without observing $H_1, \dots, H_4$, and with the dependence between $X^e$ and $H^e$ from the real data, the observed response variable does not follow a location-scale model.

\subsection{Methods}
We compare IPP on a subset of the methods from Section \ref{sec:single_cell}: we exclude distributional anchor regression, which was not competitive in any metric; for IPP, we only consider the parametric variants, which have similar performance to the neural network implementations; for point predictions, we exclude V-REx, which is computationally much more intense than but did not achieve superior performance than anchor regression and DRIG; and for conformal prediction, we only consider training on the observational environment, which was slightly superior in terms of worst case coverage on the real single cell data compared to pooling the training environments.

\subsection{Results}
\begin{figure}
    \includegraphics[width=\textwidth]{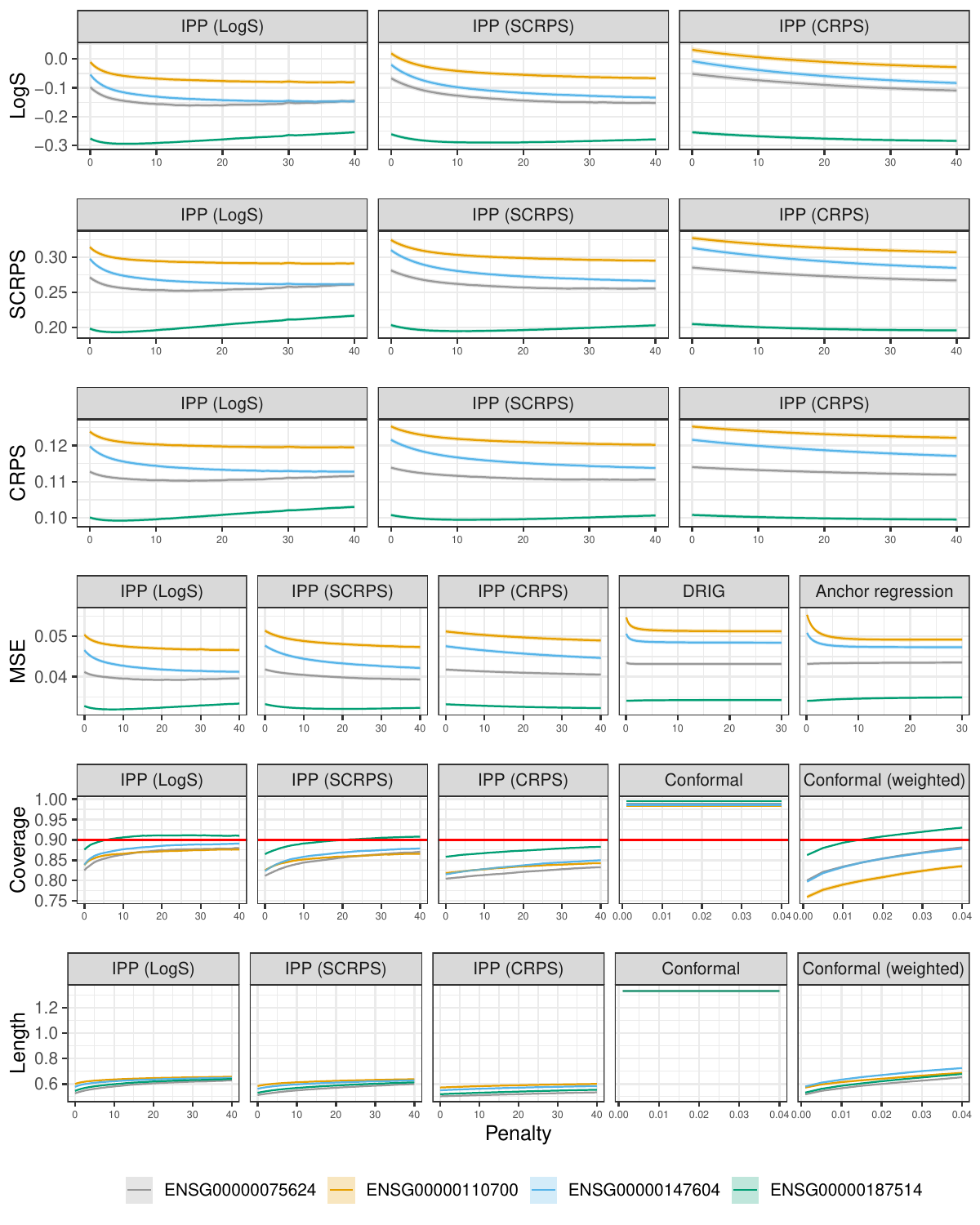}
    \caption{Evaluation metrics for the Simulations of \ref{sec:sim2}, on the environments with interventions on hidden genes. The lines are averages over $500$ simulations, and bands, which are very narrow and not visible in most cases, are delimited by mean plus/minus standard error. \label{fig:simulation2hidden}}
\end{figure}

Figure \ref{fig:simulation2hidden} compares the prediction methods on the environments with interventions on the $4$ hidden gene. Like on the real single cell data, the variants of IPP outperform anchor regression and DRIG in terms of mean squared error, and, with the exception of IPP trained with CRPS, also conformal prediction in terms of prediction interval coverage and length. For sufficiently strong penalization, IPP trained with LogS and SCRPS achieve a prediction interval coverage close to the nominal level of $90\%$ for all environments. The CRPS variant of IPP and the conformal prediction methods might benefit from stronger penalization than the maximum considered in this experiment; however, the conformal prediction intervals are already longer than those of IPP while having lower coverage. For the LogS, SCRPS, and CRPS, training IPP on LogS achieves the lowest errors, confirming the results of the real data application.

\begin{figure}
    \includegraphics[width=\textwidth]{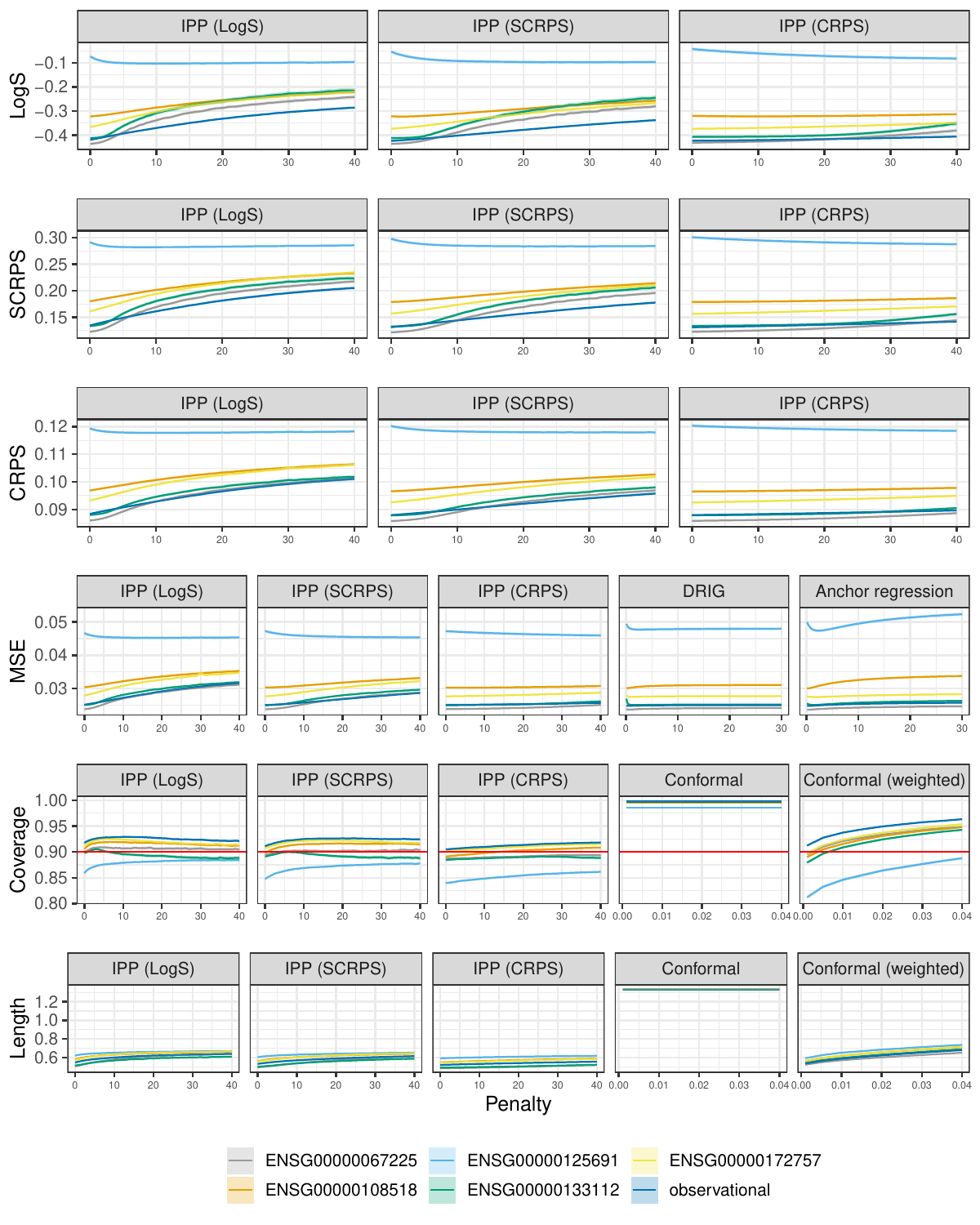}
    \caption{Like Figure \ref{fig:simulation2hidden} but for the observational environment and those with interventions on observed genes. \label{fig:simulation2obs}}
\end{figure}

In Figure \ref{fig:simulation2obs} we compare the methods in the observational environment and in those with interventions on observed genes. The picture is slightly different from the interventions on hidden genes: while IPP still achieves the smallest risk and best coverage in the worst of the environments, it does so at the cost of increasing prediction error in the other environments. This is due to the fact that IPP tries to equalize the prediction error across environments. Hence, if the interventions on test data are similar to those in the training data where the error of IPP increases with penalization, one obviously has to expect a diminished prediction accuracy. This effect also occurs in some environments for anchor regression and DRIG, but it is less pronounced. For conformal prediction the coverage in the worst environment is $80\%$ with low penalization, and increases to $90\%$ as the penalization increases, but at the cost of longer prediction intervals and overcoverage in the other environments.

To summarize, the simulations in this section are consistent with the empirical findings of Section \ref{sec:single_cell}, where IPP is consistent with or outperforms robust methods for point prediction and uncertainty quantification. As for any robust prediction method, one must bear in mind that decreasing the worst-case prediction error comes at a cost, namely, increasing the prediction error in situations where it would have been low or acceptable without any penalization towards invariance.

\end{document}